\documentclass[twocolumn,nofootinbib]{revtex4}

\usepackage{blindtext,hyperref,mathtools,amsmath,amsfonts,amssymb,physics,slashed,esvect,mathrsfs,dsfont,bbold}
\usepackage[utf8]{inputenc}
\usepackage{graphicx}
\usepackage{float}
\usepackage[footnotesize]{caption}

\usepackage{subfig}

\usepackage{multirow}

\usepackage{empheq}

\DeclareCaptionJustification{justified}{\leftskip=0pt \rightskip=0pt \parfillskip=0pt plus 1fil}

\usepackage{xcolor}

\mathchardef\Re="023C 
\mathchardef\Im="023D

\pdfoutput=1

\usepackage{environ}
\NewEnviron{myequation}{
\begin{equation}
\BODY
\end{equation}
}

\begin{document}
\title{Reconstruction of the neutron star equation of state from $w$-quasinormal modes spectra with a piecewise polytropic meshing and refinement method.}
\author{Juan Mena-Fernández}
\email[]{juanmena@ucm.es}
\author{Luis Manuel González-Romero}
\email[]{mgromero@fis.ucm.es}
\affiliation{Departamento de Física Teórica, Facultad de Ciencias Físicas, Universidad Complutense de Madrid, 28040 Madrid, Spain}

\date{\today}

\begin{abstract}
In this paper we present a new approach to the inverse problem for relativistic stars using quasinormal modes and the piecewise polytropic parametrization of the equation of state. The algorithm is a piecewise polytropic meshing and refinement method that reconstructs the neutron star equation of state from experimental data of the mass and the $wI$-quasinormal modes. We present an algorithm able to numerically calculate axial quasinormal modes of neutron stars in an efficient way. We use an initial mesh of $27440$ equations of state in a $4$-volume of piecewise polytropic parameters that contains most of the candidate equations of state used today. The refinement process drives us to the reconstruction of the equation of state with a certain precision. Using the reconstructed  equation of state, we calculate predictions for tidal deformability and slow rotation parameters (moment of inertia and quadrupole moment, for example).

In order to check the method with an explicit example, we use as input data a few (five) configurations of a given equation of state. We reconstruct the equation of state in a quite good approximation, and then we compare the curves of physical parameters from the original equation of state and the reconstructed one. We obtain a relative difference for all of the parameters smaller than $2.5\%$ except for the tidal deformability, for which we obtain a relative difference smaller than $6.5\%$. We also study constraints from GW170817 event for the reconstructed equation of state. 
\end{abstract}

\pacs{1}

\maketitle

\section{Introduction}

The detection of gravitational waves by the LIGO-VIRGO collaboration (GW150914-\cite{abbott2016observation}, GW170814-\cite{abbott2017gw170814}, GW170817-\cite{abbott2017gw170817}) opens a new era in relativistic astrophysics. In particular, the GW170817 event, which seems to be the consequence of the merging and colliding of a pair of neutron stars, can be used to study the properties of these relativistic stars. The remnant of such a process after the merging will oscillate in the ringdown phase radiating gravitational waves. These gravitational waves could be detected in future enhancements in the sensitivity of the LIGO-VIRGO detectors. The ringdown phase can be described by quasinormal modes (QNMs). In the case that the remnant was a relativistic star, the data of those expected detections could be used to obtain information about the behavior of matter in the inner part of the star, i.e., to obtain information about the equation of state (EOS).

The problem to obtain the EOS for neutron stars from macroscopic data of these stars has been treated by Lindblom using the mass-radio curve in \cite{lindblom1992determining}, recent modifications can be found in (\cite{lindblom2012spectral},\cite{lindblom2014spectral},\cite{lindblom2014relativistic}). Other authors have studied the same problem using different techniques (\cite{kokkotas2001inverse},\cite{abdelsalhin2018solving}). This problem receives generally the name of inverse stellar structure problem. In this paper we develop a method to reconstruct the EOS of neutron stars from experimental measurements of the mass and the frequency of the fundamental $wI$ mode ($wI$-QNM) of different neutron stars, i.e., from the mass-frequency $wI$-QNM curve.

The study of QNMs of neutron stars has a long tradition (\cite{regge1957stability},\cite{thorne1967non},\cite{zerilli1970gravitational}), and a summary can be found in \cite{kokkotas1999quasi}. Some recent results can be found in (\cite{blazquez2013phenomenological},\cite{blazquez2014polar}). In these papers, an efficient method to calculate the QNM spectra was presented. This method was also applied to study QNMs in alternative theories (\cite{blazquez2016axial},\cite{blazquez2018axial},\cite{blazquez2018quasinormal},\cite{motahar2018axial},\cite{blazquez2018axial_2}). Here we will present an enhancement of the method that will allow us to calculate the $w$-QNM spectra of thousands of EOSs in a reasonable time.

In Sec. \ref{overview} we briefly summarize the necessary theoretical background of the QNM approximation, starting with static and spherically symmetric stars in order to introduce axial perturbations. In Sec. \ref{numericalQNM} we explain the method developed to calculate quasinormal modes. In Sec. \ref{code} we check the numerical codes we developed by calculating macroscopic parameters for some different realistic EOSs. In Sec. \ref{ppEOS} we briefly introduce the piecewise polytropic parametrization of the EOS in order to develop our approach to the inverse problem in Sec. \ref{problema_inverso}. In sections \ref{resultadosII} and \ref{problema_inverso_resultados} we show the results of this inverse method. Finally, in Sec. \ref{conclusions} we finish the paper with a summary of the main results. In the Appendices \ref{appendixA} and \ref{appendixB} we summarize the slow rotation approximation and the calculation of the tidal Love parameter, respectively.

\section{Overview of the formalism}\label{overview}

We will here obtain the necessary differential equations to calculate the QNMs of non-rotating neutron stars. We will start with static and spherically symmetric relativistic stars and then we will introduce axial perturbations.

\subsection{Static and spherically symmetric relativistic stars}

Coordinates can be chosen so that the line element has the form
\begin{myequation}\label{sim_esf}
	ds^2=-e^{\nu(r)}(cdt)^2+e^{\lambda(r)}dr^2+r^2(d\theta^2+\sin^2\theta d\varphi^2).
\end{myequation}
We will consider the matter in the interior of the star as an effective perfect fluid with a barotropic equation of state.
$u^\mu$ is the fluid's $4$-velocity, $p$ is the pressure and $\epsilon$ is the energy density$/c^2$. 

It is widely known that the equations describing static and spherically symmetric relativistic stars are given by
\begin{subequations}\label{orden0}
	\begin{gather}
	\begin{align}
	&\frac{dm}{dr}=4\pi r^2\epsilon,\\
	&\frac{d\nu}{dr}=\frac{2G}{c^2r^2}\frac{m+\frac{4\pi}{c^2}r^3p}{1-\frac{2Gm}{c^2r}},\\
	&\frac{dp}{dr}=-\left(\epsilon+\frac{p}{c^2}\right)\frac{G}{r^2}\frac{m+\frac{4\pi}{c^2}r^3p}{1-\frac{2Gm}{c^2r}},
	\end{align}
	\end{gather}
\end{subequations}
where
\begin{myequation}
	m=\frac{c^2r}{2G}(1-e^{-\lambda}).
\end{myequation}
Provided an equation of state, $p=p(\epsilon)$, the system of ordinary differential equations \eqref{orden0} can be solved numerically.

\subsubsection{Exterior solution for non-rotating stars}

The exterior spacetime is given by the Schwarzschild solution, 
\begin{myequation}
	\nu=-\lambda=\log\left(1-\frac{2GM}{c^2r}\right),
\end{myequation}
where $M$ is the mass of the star.

From the junction conditions between the interior and the exterior solutions it follows that $p(R)=0$, with $R$ the radius of the star.

\subsection{Quasinormal modes of non-rotating stars}

Pulsating stars are very important sources of information for astrophysics. Almost every star pulsates during its evolution from the early stages to the very late ones, when the catastrophic creation of a compact object  (white dwarf, neutron star or black hole) occurs. Non-radial pulsations of compact objects are accompanied with gravitational wave emission \cite{kokkotas1999quasi}.

Gravitational waves are oscillations of spacetime, typically produced by matter oscillations, which propagate throughout empty spacetime. This is a result of  General Relativity, since spacetime has its own dynamics and it is coupled to matter via the Einstein field equations. The possible sources of gravitational waves are interacting black holes, coalescing compact binary systems, stellar collapses and pulsars. Thus, neutron stars are good candidates as sources of detectable gravitational radiation. In fact, the gravitational waves emmited by a neutron star inspiral were first observed in August of 2017 (GW170817-\cite{abbott2017gw170817}).

It is well known that neutron star's characteristic oscillations present a frequency and a decay time (damping time). The spectrum of resonances of these gravitational waves can give us information about the structure and composition of neutron stars, i.e., information about the EOS of neutron stars.

The appropriate mathematical tool to study these damped oscillations is the quasinormal mode (QNM) expansion. Mathematically, QNM expansion is implemented via perturbations over static and spherically symmetric spacetime. If we assume that the pulsation of the star is small, we can use perturbation theory to describe the oscillation,
\begin{myequation}
	g_{\mu\nu}=g_{\mu\nu}^{(0)}+h_{\mu\nu}+\mathcal{O}(\chi^2),
\end{myequation}
where $g_{\mu\nu}^{(0)}$ is given by Eq. \eqref{sim_esf} and $h_{\mu\nu}=\mathcal{O}(\chi)$. The QNM expansion parameter has been denoted as $\chi$. We will keep terms up to first order in the perturbation. Matter is also perturbed, as well as the fluid's $4$-velocity,
\begin{subequations}
	\begin{gather}
	\epsilon\to\epsilon+\delta\epsilon+\mathcal{O}(\chi^2),\\
	p\to p+\delta p+\mathcal{O}(\chi^2),\\
	u\to u+\frac{\partial\xi}{\partial t}+\mathcal{O}(\chi^2),
	\end{gather}
\end{subequations}
where $\xi$ is the displacement of a fluid element.

In general, for non-radial oscillations, the perturbation functions are dependent of radial and angular coordinates, and time. The usual procedure is to expand the angular dependence of the functions in spherical harmonics. Under a parity transformation, $\vec{x}\to-\vec{x}$ ($\theta\to \theta+\pi$ in spherical coordinates), the components of the metric and stress-energy tensor will transform like scalars, vectors or higher type tensors. Perturbations are, then, indexed with integer numbers $(l,m)$, but axial symmetry will remove the dependency on the $m$ number.

There are three type of perturbations: radial, axial and polar perturbations. Since we are considering solutions of General Relativity, the radial perturbations are not interesting for gravitational wave search, so they won't be studied here. Under a parity transformation, axial perturbations transform like $(-1)^{l+1}$, while polar ones transform like $(-1)^l$. Hence, axial perturbations do not couple to polar ones, so we can consider them separately. In this work, we will only consider axial perturbations.

\subsubsection{Axial perturbations}

Here we will obtain the so-called Regge-Wheeler equation \cite{regge1957stability}. In the Regge-Wheeler gauge, axial perturbations are described by the metric perturbation 
\begin{widetext}
\begin{myequation}
h_{\mu\nu}^{(axial)}=\sum_{l,m}
\begin{bmatrix}
0 & 0 & -h_0^{l,m}\frac{1}{\sin\theta}\frac{\partial Y_{lm}}{\partial\varphi} & h_0^{l,m}\sin\theta\frac{\partial Y_{lm}}{\partial\theta}\\
0 & 0 & -h_1^{l,m}\frac{1}{\sin\theta}\frac{\partial Y_{lm}}{\partial\varphi} & h_1^{l,m}\sin\theta\frac{\partial Y_{lm}}{\partial\theta}\\
-h_0^{l,m}\frac{1}{\sin\theta}\frac{\partial Y_{lm}}{\partial\varphi} & -h_1^{l,m}\frac{1}{\sin\theta}\frac{\partial Y_{lm}}{\partial\varphi} & 0 & 0\\
h_0^{l,m}\sin\theta\frac{\partial Y_{lm}}{\partial\theta} & h_1^{l,m}\sin\theta\frac{\partial Y_{lm}}{\partial\theta} & 0 & 0
\end{bmatrix},
\end{myequation}
\end{widetext}
where $h_0$ and $h_1$ are functions of $t$ and $r$, and first order in $\chi$. It is possible to choose the gauge such that the perturbation of the $4$-velocity is trivial,
\begin{myequation}
\delta u^{(axial)\mu}=0.
\end{myequation}
As we already mentioned, axial symmetry simplifies these expressions. It is well known that
\begin{myequation}
Y_{lm}(\theta,\varphi)\propto P_l^m(\cos\theta)e^{im\varphi},\nonumber
\end{myequation}
and so we can write $h_{\mu\nu}^{(axial)}$ in terms of the associated Legendre polynomials, $P_l^m(\cos\theta)$. Because of axial symmetry, we can set $m=0$. The associated Legendre polynomials are related to the non-associated ones by $P_l^0(\cos\theta)=P_l(\cos\theta)$. Hence,
\begin{myequation}
h_{\mu\nu}^{(axial)}=\sum_{l}\begin{bmatrix}
0 & 0 & 0 & h_0^l\\
0 & 0 & 0 & h_1^l\\
0 & 0 & 0 & 0\\
h_0^l & h_1^l& 0 & 0
\end{bmatrix}\sin\theta\frac{\partial P_l(\cos\theta)}{\partial\theta}.
\end{myequation} 
Different values of $l$ give rise to different equations decoupled one from each other. From now on, only $l=2$ perturbations will be considered.

Instead of $h_1$, it is easier to write the Einstein field equations for a metric function $Z$ defined as
\begin{myequation}
	Z=e^{\frac{\nu-\lambda}{2}}\frac{h_1}{r}.
\end{myequation}
The resulting equation of motion for $Z$ is
\begin{myequation}
	\frac{\partial^2Z}{c^2\partial t^2}=e^{\nu-\lambda}\frac{\partial^2Z}{\partial r^2}+e^{\nu-\lambda}\left\{-\left[\frac{1}{r}\left(1-e^\lambda\right)+\frac{4\pi G}{c^2}re^\lambda\right.\right.\nonumber
\end{myequation}
\vspace{-0.4cm}
\begin{myequation}\label{Regge-Wheeler}
	\left.\left.\times \left(\epsilon-\frac{p}{c^2}\right)\vphantom{\frac{1}{r}}\right]\frac{\partial Z}{\partial r}-\left[\frac{3}{r^2}\left(1+e^\lambda\right)+\frac{4\pi G}{c^2}e^\lambda\left(\epsilon-\frac{p}{c^2}\right)\right]Z\vphantom{\frac{1}{r}}\right\}.
\end{myequation}

A change of variable to the so-called \textit{tortoise coordinate} \cite{kokkotas1999quasi}, $dr_*=e^{\frac{\lambda-\nu}{2}}dr$, turns the previous expression into a wave equation with a potential barrier,
\begin{myequation}
	\frac{\partial^2Z}{c^2\partial t^2}-\frac{\partial^2Z}{\partial r_*^2}+e^\nu\left[\frac{6}{r^2}\left(1-\frac{Gm}{c^2r}\right)+\frac{4\pi G}{c^2}\left(\epsilon-\frac{p}{c^2}\right)\right]Z=0.
\end{myequation}
This equation is commonly known as the Regge-Wheeler equation.

\subsubsection{The quasinormal modes of a star}\label{QNMsteps}

The quasinormal modes are solutions to the axial (and polar) equations that satisfy the following boundary conditions:
\begin{enumerate}
	\item All perturbed functions (in our case, $Z$) have a regular behavior at $r=0$.
	\item They behave as a pure outgoing wave at infinity.
	\item The interior solution matches with the exterior one at the surface of the star.
\end{enumerate}

The QNM spectrum of a star in General Relativity has a very rich structure. We have already seen that the equations governing the axial perturbations can be reduced to a single wave equation with a potential barrier for the metric function $Z$, Eq. \eqref{Regge-Wheeler}. The role of the fluid is that of determining the shape of the potential barrier, as it depends on $m$, $\epsilon$ and $p$ \cite{husa2003current}. Hence, the axial QNMs are pure gravitational modes and do not have a newtonian counterpart. This is the reason why axial QNMs are known as spacetime modes (or $w$ modes).

On the other hand, the polar modes (which won't be numerically calculated here) are essentially fluid pulsations, and are known as fluid modes. The classification of polar modes can be found in Ref. \cite{kokkotas1999quasi}.

As we have already mentioned, $w$ modes are spacetime modes, and hence they don't have a newtonian counterpart. These modes exist for both axial and polar oscillations. In Sec. \ref{numericalQNM}, we will develop a numerical scheme to obtain these $w$ modes. In particular, we will focus on the calculation of the fundamental $wI$ mode.

\section{Numerical calculation of quasinormal modes}\label{numericalQNM}

In this section we will explain the algorithm we developed to calculate QNMs of neutron stars. The equation to be solved is the Regge-Wheeler equation, Eq. \eqref{Regge-Wheeler}. The time dependence of the metric function $Z$ will be written as
\begin{myequation}
	Z(t,r)=e^{-i\omega t}Z(r).
\end{myequation}
The complex frequency $\omega$ will be expressed as
\begin{myequation}
	\omega=\Re\omega+i\Im\omega=2\pi\nu+i\frac{1}{\tau},
\end{myequation}
where $\nu$ is the frequency of the oscillation and $\tau$ is the damping time. In the interior of the star we have the equation
\begin{myequation}
	\frac{d^2Z}{dr^2}+\left(1-\frac{2Gm}{c^2r}\right)^{-1}\left\{-\left[-\frac{2Gm}{c^2r^2}+\frac{4\pi G}{c^2}r\left(\epsilon-\frac{p}{c^2}\right)\right]\frac{dZ}{dr}\right.\nonumber
\end{myequation}
\vspace{-0.3cm}
\begin{myequation}
	\left.-\left[\frac{6}{r^2}\left(1-\frac{Gm}{c^2r}\right)+\frac{4\pi G}{c^2}\left(\epsilon-\frac{p}{c^2}\right)\right]Z+\frac{\omega^2}{c^2}e^{-\nu}Z\right\}=0.
\end{myequation}

On the other hand, in the exterior of the star we have the equation
\begin{myequation}
	\frac{d^2Z}{dr^2}+\left(1-\frac{2GM}{c^2r}\right)^{-1}\left[\frac{2GM}{c^2r^2}\frac{dZ}{dr}\qquad\qquad\qquad\qquad\right.\nonumber
\end{myequation}
\vspace{-0.5cm}
\begin{myequation}
	\left.-\frac{6}{r^2}\left(1-\frac{GM}{c^2r}\right)Z+\frac{\omega^2}{c^2}\left(1-\frac{2GM}{c^2r}\right)^{-1}Z\right]=0.
\end{myequation}
Instead of $Z$, we will consider the phase function $g$,
\begin{myequation}
	g=\frac{1}{Z}\frac{dZ}{dr},
\end{myequation}
which doesn't oscillate towards asymptotic infinity \cite{blazquez2013phenomenological}. The equation for $g$ is given by
\begin{myequation}\label{eq10}
	\frac{dg}{dr}+g^2+\left(1-\frac{2GM}{c^2r}\right)^{-1}\left[\frac{2GM}{c^2r^2}g\right.\qquad\qquad\qquad\qquad\nonumber
\end{myequation}
\vspace{-0.3cm}
\begin{myequation}
	\left.-\frac{6}{r^2}\left(1-\frac{GM}{c^2r}\right)+\frac{\omega^2}{c^2}\left(1-\frac{2GM}{c^2r}\right)^{-1}\right]=0.
\end{myequation}

To develop the numerical scheme, we will follow the three steps we listed in \ref{QNMsteps}.

\vspace{0.07cm}\begin{center} $\boldsymbol{1.}$ \textbf{Interior solution: regular behavior at} $\boldsymbol{r=0}$\end{center}\vspace{0.07cm}

The metric function $Z$ has to satisfy the condition of regular behavior at $r=0$. Hence, we have to expand $Z$ in taylor series and find its coefficients at $r=0$. The asymptotic behavior of $Z$ when $r\to 0$ can be easily calculated and is given by
\begin{myequation}
	Z=a_3\left[r^3+\frac{16\pi G\left(\epsilon_0-p_0/c^2\right)-\omega^2e^{-\nu_0}}{14c^2}r^5+\mathcal{O}\left(r^7\right)\right].
\end{myequation}

\vspace{0.07cm}\begin{center} $\boldsymbol{2.}$ \textbf{Exterior solution: outgoing wave behavior}\end{center}\vspace{0.07cm}

Now we have to impose the purely outgoing wave behavior, i.e., $g\to -i\frac{\omega}{c}$ when $r\to\infty$ \cite{blazquez2013phenomenological}. A complexification of the radial coordinate is a commonly used technique to obtain solutions with no ingoing wave contamination \cite{blazquez2013phenomenological} (this technique is known as Exterior Complex Scaling \cite{nicolaides1990resonances}). The equation for $g$ is now written in terms of a new variable $y$ related to $r$ by
\begin{myequation}
	r=R+ye^{i\alpha},\quad y\in [0,\infty).
\end{myequation}
The parameter $\alpha$ must satisfy
\begin{myequation}
	\Re\omega\sin\alpha+\Im\omega\cos\alpha<0,
\end{myequation}
see Ref. \cite{blazquez2016axial}. In order to impose the outgoing wave behavior condition at spatial infinity as accurately as possible, it is necessary to compactify $y$. We found useful to compactify it as
\begin{myequation}
	y=\frac{1-x}{x},\quad x\in[1,0).
\end{myequation}
Now we have a compactified complex radial coordinate,
\begin{myequation}
	r=R+\frac{1-x}{x}e^{i\alpha},\quad x\in[1,0).
\end{myequation}

The asymptotic behavior of $g(x)$ when $x\to 0$ ($\sim r\to\infty$) can be easily calculated. It is, for an outgoing wave, given by
\begin{myequation}
	g=-i\frac{\omega}{c}\left[1+\frac{2GM}{c^2}e^{-i\alpha}x+\mathcal{O}(x^2)\right].
\end{myequation}

\vspace{0.07cm}\begin{center} $\boldsymbol{3.}$ \textbf{Matching the interior and exterior solutions}\end{center}\vspace{0.07cm}

Now we have to impose the third condition: the interior solution has to match with the exterior one at the surface of the star. The junction condition between the interior solution and the exterior one on the boundary of the star is given by the continuity of the phase function $g$ \cite{blazquez2013phenomenological}, 
\begin{myequation}\label{junction}
	g^{(\omega)}_{in}(R)=g^{(\omega)}_{out}(R),
\end{myequation}
where \textit{in} indicates inside the star and \textit{out} denotes out of it. The algorithm implemented to find the quasinormal modes for a given EOS is a Müller-type method, taken from Ref. \cite{press1996numerical}. The function whose roots have to be found to obtain $\omega$ is $f(\omega)$, which is defined as
\begin{myequation}
	f(\omega)\equiv g_{in}^{(\omega)}(R)-g_{out}^{(\omega)}(R).
\end{myequation}

\begin{figure}[H]
	\centering
	\includegraphics[width=\linewidth]{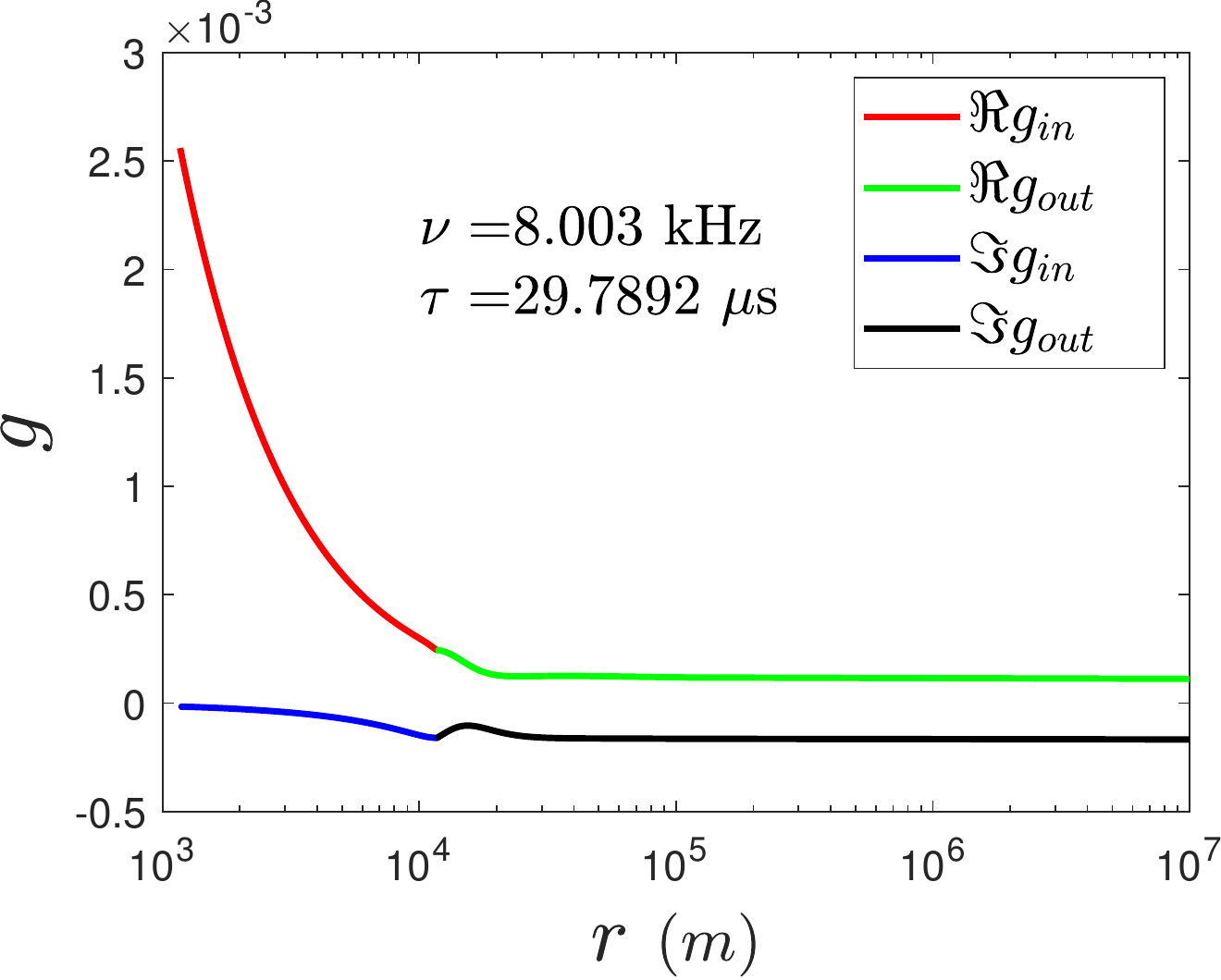}\captionsetup{width=0.5\textwidth}\caption{Real and imaginary part of the phase function $g$ vs $r$ inside and outside the star for a fundamental $wI$ mode with $\nu=8.003$ kHz and $\tau=29.7892$ $\mu$s, obtained for SLy EOS with central energy density $\epsilon_0=10^{18}$kg/m${}^3$ ($M=1.4169M_\odot$).}\label{fig1}
\end{figure}

We will focus on the numerical calculation of the fundamental $wI$ mode. In Fig. \ref{fig1} we show an example of a fundamental $wI$ mode for SLy EOS.

\section{The code analysis}\label{code}

We have implemented the numerical method explained in Sec. \ref{numericalQNM} to calculate the quasinormal modes in a PYTHON code with some parallelization. In order to check the code we have used well known equations of state of different types (EOS with plain nuclear matter, with hyperons, for hybrids stars and for quark stars). We list below the different models of EOSs used to check the code.

\begin{itemize}
	\item For plain $npe\mu$ nuclear matter we use
	\begin{itemize}
		\item APR4 EOS \cite{akmal1998equation}, obtained using a variational method.
		\item SLy EOS \cite{douchin2001unified}, obtained using a potential-method.
	\end{itemize}
	\item For mixed hyperon-nuclear mater we use
	\begin{itemize}
		\item GNH3 EOS \cite{glendenning1984neutron}, a relativistic mean-field theory EOS containing hyperons.
		\item BHZBM EOS \cite{bednarek2012hyperons}, a non-linear relativistic mean field model involving baryon octet coupled to meson fields.
	\end{itemize}
	\item For hybrid stars we use ALF4 EOS \cite{alford2005hybrid}, a hybrid EOS with mixed APR nuclear matter and color-flavor-locked quark matter.
	\item For hybrid stars with hyperons and quark color-superconductivity we use BS3 EOS \cite{bonanno2012composition}, obtained using a combination of phenomenological relativistic hyper-nuclear density functional and an effective NJL model of quantum chromodynamics. The parameters considered are vector coupling $G_V/G_S=0.6$ and quark-hadron transition density $\rho_{tr}/\rho_0=3.5$, where $\rho_0$ is the density of nuclear saturation.
	\item For quark stars we use WSPHS EOS \cite{weissenborn2011quark}, an unpaired quark matter EOS with parameters $B_{eff}^{1/4}=123.7$ MeV and $a_4=0.53$.
\end{itemize}

\begin{figure}[H]
\centering
\includegraphics[width=\linewidth]{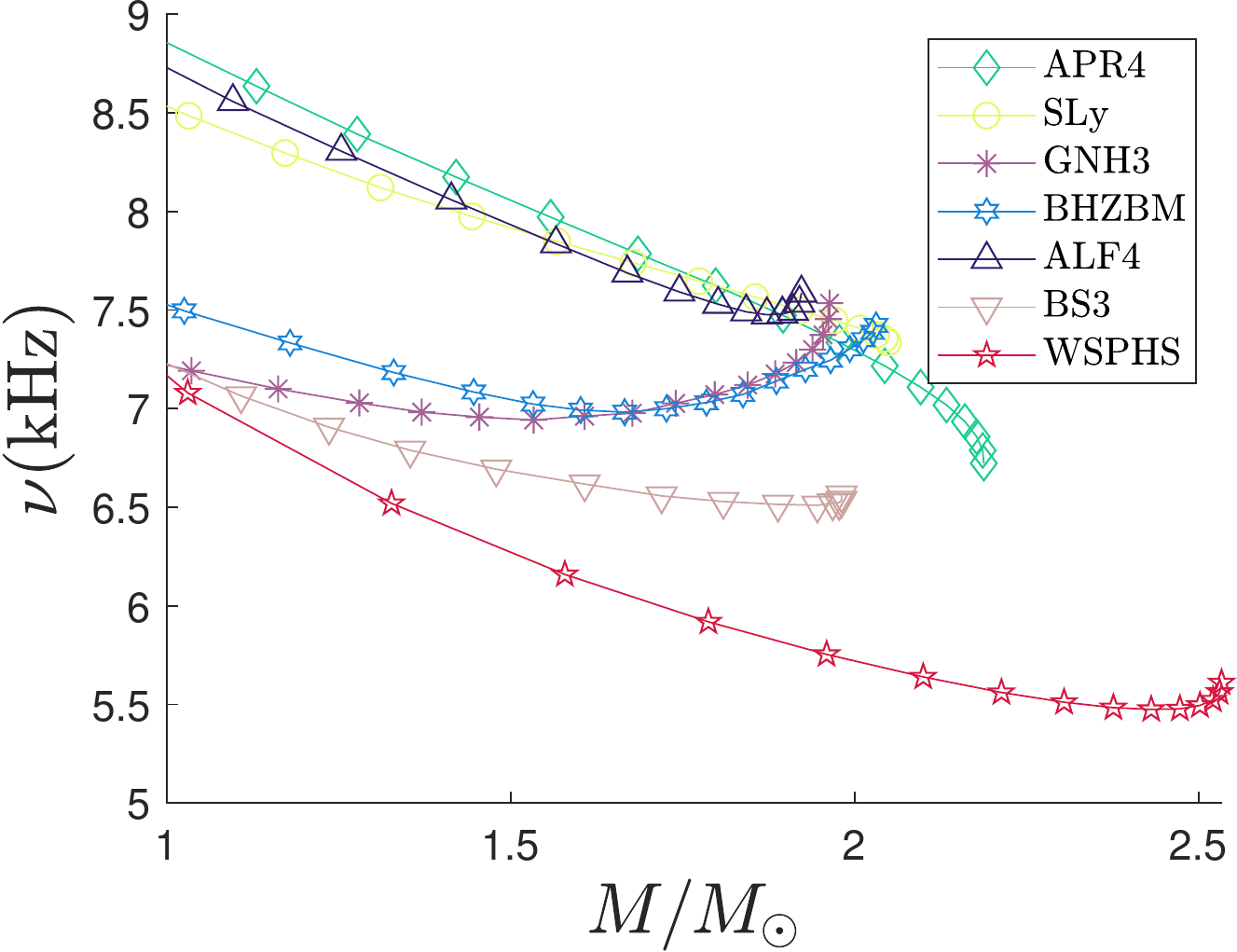}\captionsetup{width=0.5\textwidth}\caption{Frequency of the fundamental $wI$ mode vs mass for the different EOSs considered.}\label{fig2}
\end{figure}

The results of applying the code to these models can be found in Fig. \ref{fig2}. In this figure we present the frequency of the fundamental $wI$ mode versus the mass ($M-\nu$) for the different EOSs considered, with $20$ configurations for each EOS. In our computer it takes about $90$ seconds to calculate the $20$ quasinormal modes for one EOS. The results are compatible with those shown in Ref. \cite{blazquez2013phenomenological}. 

For our method to reconstruct the EOS of neutron stars, it will be necessary to calculate thousands of these $M-\nu$ curves, as will be explained in Sec. \ref{problema_inverso}.

Once the EOS is reconstructed, we will be able to make predictions of other macroscopic parameters. We will consider  parameters calculated for slowly rotating relativistic stars, together with the tidal deformability (tidal Love parameter). In the Appendices \ref{appendixA} and \ref{appendixB} we present a summary of the equations and definitions used. We have also implemented a Python code to calculate the global parameters obtained in these Appendices (moment of inertia, quadrupole momment, tidal Love parameter,$\dots$). Using the results it is possible to check the compatibility of the equations of state with the observations of the gravitational waves in the GW170817 event. To check the code that we will use later, we will apply it to the EOSs considered in this section. 

\begin{figure}[H]
\centering
\includegraphics[width=\linewidth]{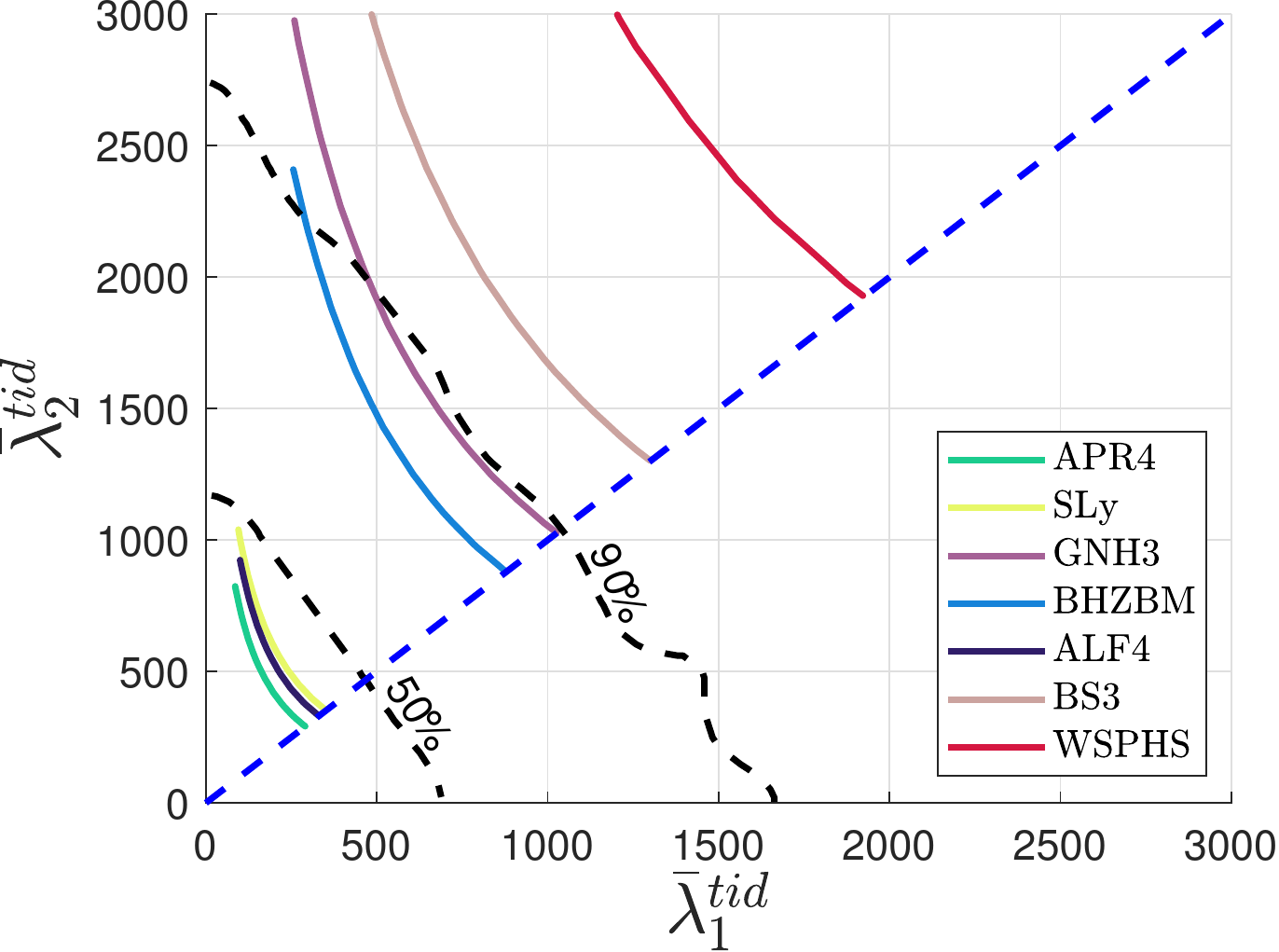}\captionsetup{width=0.5\textwidth}\caption{Predictions for tidal deformability given by the different realistic EOSs considered, under the assumption that both components are neutron stars (for the low-spin scenario). Contours enclosing $90\%$ and $50\%$ of the probability density are shown as dashed lines (both curves taken from Ref. \cite{abbott2017gw170817}).}\label{fig3}
\end{figure}

The observation of gravitational waves provides new information about which models of EOSs are more likely to be candidate EOSs. GW170817 event was the first observation of gravitational waves from a binary neutron star inspiral \cite{abbott2017gw170817}. In Ref. \cite{abbott2017gw170817}, the probability density for the distribution of the tidal Love parameters of the two stellar components, $p\left[\bar{\lambda}_1^{tid},\bar{\lambda}_2^{tid}\right]$, is obtained using a post-newtonian model. Here, we will calculate the relation $\bar{\lambda}_2^{tid}(\bar{\lambda}_1^{tid})$ for the different candidate EOSs considered in this paper for a chirp mass $\mathcal{M}=(1.188^{+0.004}_{-0.002}) M_\odot$ \cite{abbott2017gw170817}, where
\begin{myequation}
	\mathcal{M}=\frac{(M_1M_2)^{3/5}}{(M_1+M_2)^{1/5}}.
\end{myequation}
The results are shown in Fig. \ref{fig3} (for the low-spin scenario), together with the contours enclosing $90\%$ and $50\%$ of the probability density (curves taken from Fig. 5 of Ref. \cite{abbott2017gw170817}). BS3 EOS and WSPHS EOS predict $\bar{\lambda}^{tid}$ values outside the $90\%$ confidence contour.

Our code reproduces previously known results for the tidal deformability. For instance, the results shown in Fig. \ref{fig3} for APR4 and SLy coincide with those presented in Ref. \cite{abbott2017gw170817}.

\section{Piecewise polytropic equations of state}\label{ppEOS}

An equation of state is polytropic if
\begin{myequation}
	p(\rho)/c^2=K\rho^\Gamma,
\end{myequation}
where $\rho$ is the rest mass density and $\Gamma$ is the polytropic index, which is defined as
\begin{myequation}
	\Gamma=\frac{\rho}{p}\frac{dp}{d\rho}.
\end{myequation}
The energy density$/c^2$, $\epsilon$, is given in terms of $\rho$ and $p$ via the first law of thermodynamics,
\begin{myequation}
	\frac{d\epsilon}{d\rho}=\frac{\epsilon+p/c^2}{\rho}.
\end{myequation}
For the polytropic case, one finds that
\begin{myequation}
	\epsilon(\rho)=(1+a)\rho+\frac{K}{\Gamma-1}\rho^\Gamma,
\end{myequation}
where $a$ is a constant \cite{read2009constraints}. 

Following Ref. \cite{read2009constraints}, an EOS is said to be piecewise polytropic for $\rho\geq\rho_0$ if, for a set of dividing densities $\rho_0<\rho_1<\rho_2<\dots$, the pressure and the energy density are continuous and given, in the interval $\rho_{i-1}\leq\rho\leq\rho_i$, by
\begin{myequation}
	p(\rho)/c^2=K_i\rho^{\Gamma_i}\quad\text{and}\quad \epsilon(\rho)=(1+a_i)\rho+\frac{K_i}{\Gamma_i-1}\rho^{\Gamma_i},
\end{myequation}
respectively, with
\begin{myequation}
	a_i=\frac{\epsilon(\rho_{i-1})}{\rho_{i-1}}-1-\frac{K_i}{\Gamma_i-1}\rho^{\Gamma_i-1}_{i-1}.
\end{myequation}
Continuity of pressure, $p(\rho_i)/c^2=K_i\rho_i^{\Gamma_i}=K_{i+1}\rho_i^{\Gamma_{i+1}}$, restricts $K_{i+1}$ to
\begin{myequation}
	K_{i+1}=\frac{p(\rho_i)/c^2}{\rho_i^{\Gamma_{i+1}}}.
\end{myequation}

\begin{figure}[H]
	\centering
	\includegraphics[width=\linewidth]{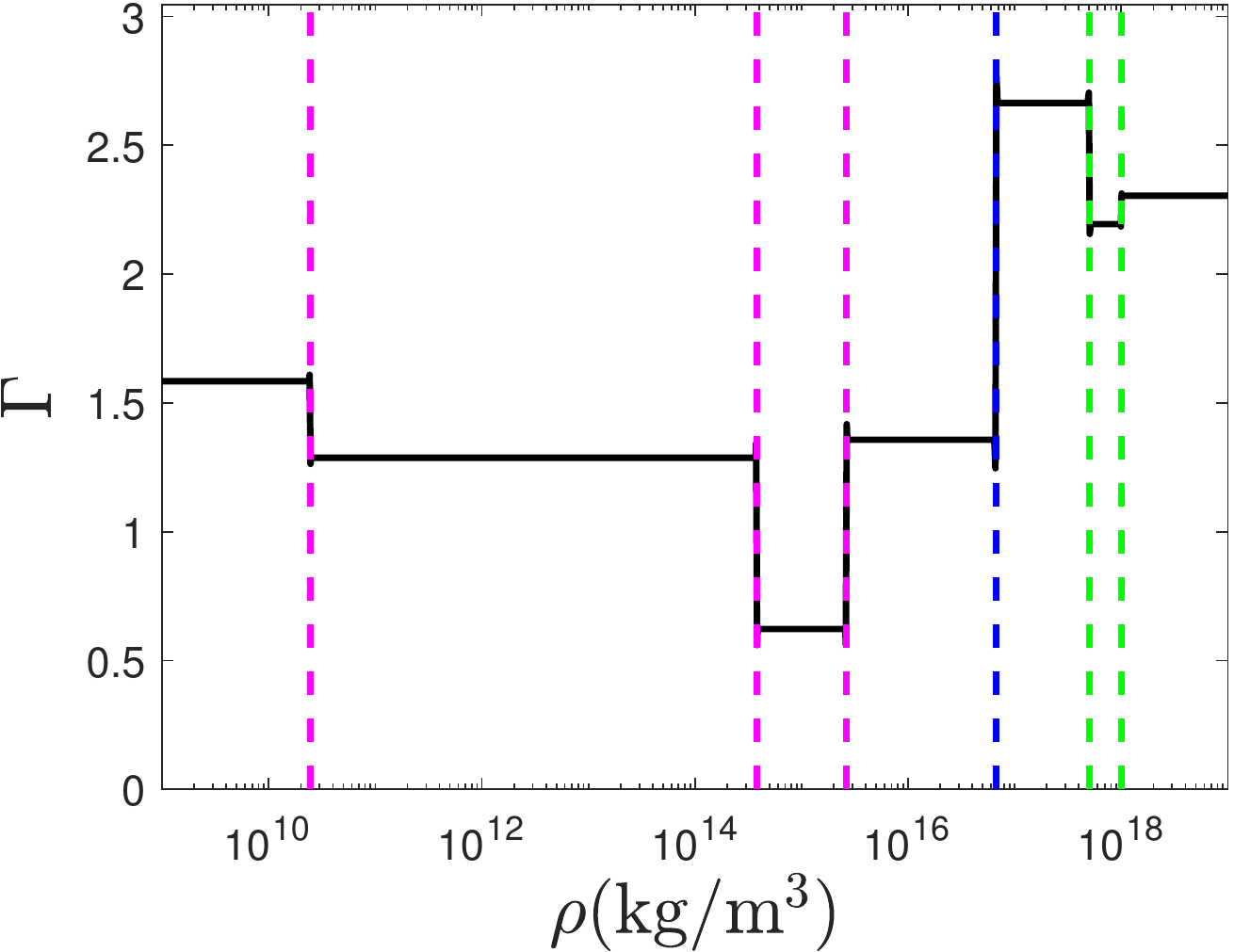}\captionsetup{width=0.5\textwidth}\caption{Reconstruction of SLy EOS from its polytropic parameters. Pink vertical lines represent fixed density values where $\Gamma$ changes in the curst region. Green vertical lines represent fixed density values where $\Gamma$ changes in the core region. Blue vertical line represents the core-crust density transition.}\label{fig4}
\end{figure}

There is general agreement on the low-density EOS for cold matter. This means that all realistic EOS have practically the same behavior below nuclear density (the so-called crust region). Hence, we can choose SLy EOS, for example, as our low-density EOS. An analytic representation of $p(\rho)$ for SLy EOS below nuclear density uses four polytropic pieces \cite{read2009constraints}. The four regions roughly correspond to a non-relativistic electron gas, a relativistic electron gas, neutron drip, and the density range from neutron drip to nuclear density \cite{read2009constraints}. The EOS below nuclear density is, then, fixed (for SLy EOS, the crust corresponds to the density region below the blue vertical line shown in Fig. \ref{fig4})

Each choice of a piecewise polytropic EOS above nuclear density (the so-called core region) is matched to this low-density EOS. In the core, we consider three polytropic pieces, specified by the six parameters $\{p_1,\Gamma_1,\rho_1,\Gamma_2,\rho_2,\Gamma_3\}$. A good fit is found for these three polytropic pieces with fixed divisions: $\rho_1=10^{17.7}$kg/m${}^3$ and $\rho_2=10^{18.0}$kg/m${}^3$, as shown in Ref. \cite{read2009constraints}. Hence, we reduce the number of parameters from six to four, namely $\{p_1,\Gamma_1,\Gamma_2,\Gamma_3\}$ (for SLy EOS, the core corresponds to the density region above the blue vertical line shown in Fig. \ref{fig4}).

At intermediate densities, the core's first polytropic piece is matched dynamically to the fixed crust, which was chosen to be a parametrized four-piece polytropic version of SLy EOS. This defines the core-crust density transition (for SLy EOS, the core-crust density transition corresponds to the blue vertical line shown in Fig. \ref{fig4}).

In Fig. \ref{fig4} we show SLy EOS reconstructed from its polytropic parameters, taken from Table III of Ref. \cite{read2009constraints}.

A wide variety of candidate and realistic EOSs, including plain nuclear matter, hyperons, condensates and deconfined quarks, are well fit by some set of polytropic parameters, namely $\{p_1,\Gamma_1,\Gamma_2,\Gamma_3\}$ \cite{read2009constraints}. Because of this, piecewise polytropic parametrization can be very useful to generate realistic EOS. This is where the idea of our approach to the inverse problem arises, as we will explain in Sec. \ref{problema_inverso}.

\section{The piecewise polytropic meshing and refinement method for the inverse problem}\label{problema_inverso}

The complete knowledge of the neutron star EOS makes possible the calculation of macroscopic quantities such as the mass, the quasinormal modes, the tidal Love parameter, etc. Viceversa, from the measurement of macroscopic observables it is possible to invert this map and reconstruct the EOS: this is the so-called \textit{inverse problem} \citep{lindblom1992determining}.

\begin{figure}[H]
	\centering
	\includegraphics[width=\linewidth]{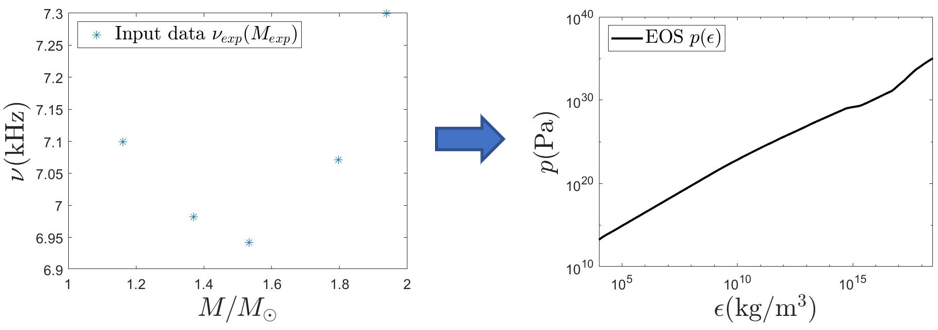}\captionsetup{width=0.5\textwidth}\caption{Illustration of the inverse problem for neutron stars. The EOS is reconstructed from measurements of the frequency of the fundamental $wI$ mode and the mass of $5$ different neutron stars.}\label{fig5}
\end{figure}

As we mentioned in Sec. \ref{ppEOS}, the piecewise polytropic parametrization fits a large class of realistic and candidate EOSs (in fact, the polytropic parameters $\{\log_{10}p_1,\Gamma_1,\Gamma_2,\Gamma_3\}$ for a wide variety of EOSs can be found in Table III of Ref. \cite{read2009constraints}). Thus, each EOS is determined simply by specifying 4 numbers. Here is where the idea of our inverse stellar method arises: we will create a mesh of EOSs in a $4$-volume of piecewise polytropic parameters $\{\log_{10}p_1,\Gamma_1,\Gamma_2,\Gamma_3\}$. This $4$-volume must include as many candidate EOSs as possible, for example, from the ones listed in Table III of Ref. \cite{read2009constraints}. The initial mesh of polytropic parameters will be given by 
\begin{myequation}\label{parametrosI}
	\begin{array}{ll}
		\log_{10}p_1=[34.2,34.7]_{10},\\\\
		\Gamma_1=[2,3.8]_{14},\\\\
		\Gamma_2=[1.8,3.8]_{14},\\\\
		\Gamma_3=[1.8,3.8]_{14}.
	\end{array}
\end{myequation}
From now on, the sub-index in an interval will indicate the number of equidistant elements taken in that interval. Hence, we will have a total of $10\times 14^3=27440$ EOSs in our initial $4$-volume, i.e., $27440$ points in a $4$-dimensional space of coordinates $\{\log_{10}p_1,\Gamma_1,\Gamma_2,\Gamma_3\}$.

As shown in the illustration of the inverse problem, Fig. \ref{fig5}, we will reconstruct the neutron star EOS from measurements of the frequency of the fundamental $wI$ mode ($\nu$) and the mass ($M$) of some different neutron stars. The input data will be denoted as $M_{\text{exp}}$ and $\nu_{\text{exp}}$.

Our algorithm will numerically calculate each $\nu_i(M_i)$ curve ($i=1,\dots,27440$) in order to find the most similar $\nu(M)$ curve to the input data $\nu_{\text{exp}}(M_{\text{exp}})$. That is, it will find the point in the $4$-space of coordinates $\{\log_{10}p_1,\Gamma_1,\Gamma_2,\Gamma_3\}$ that represents the input data with a certain precision.

\begin{figure}[H]
\centering
\includegraphics[width=0.53\linewidth]{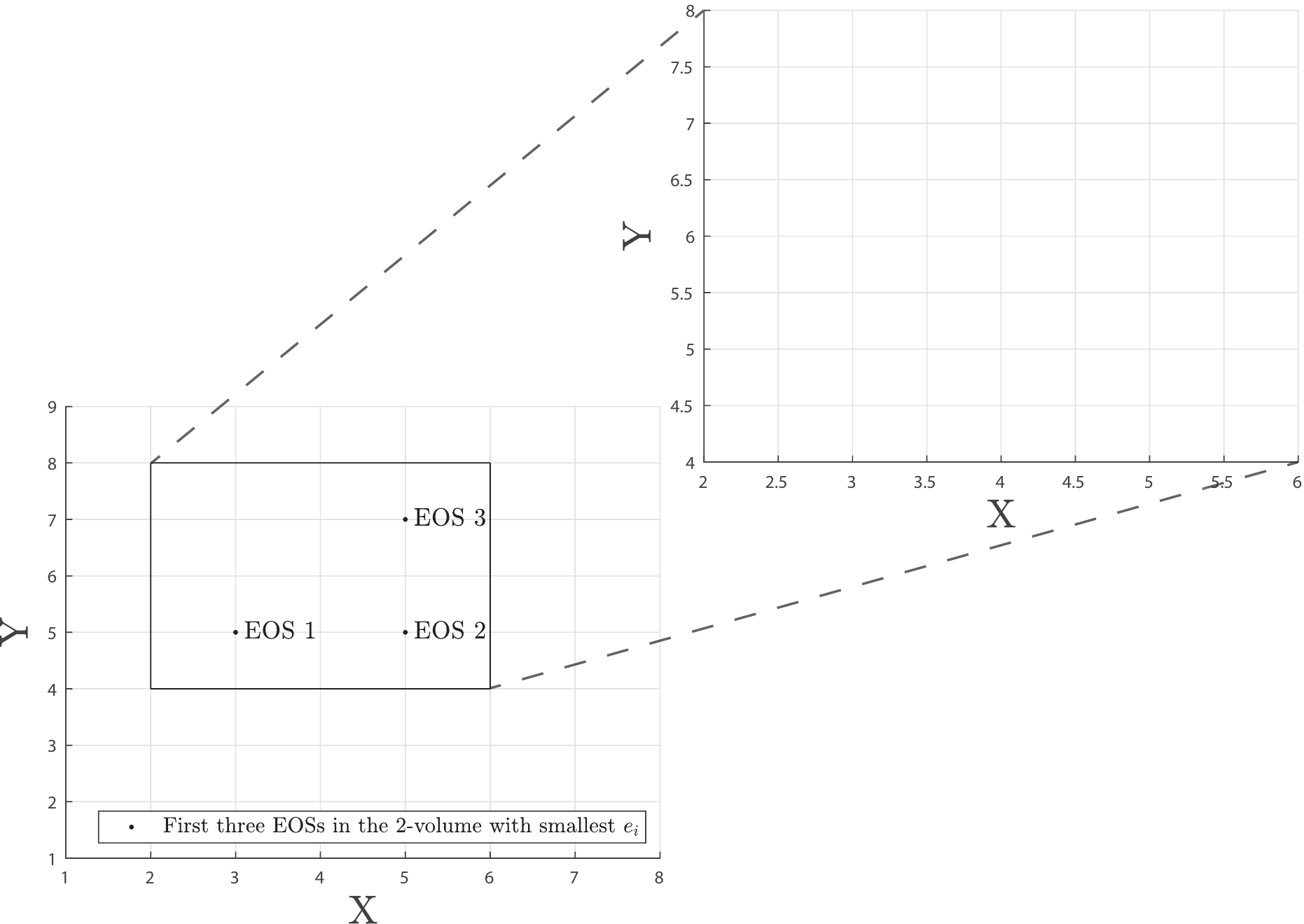}\captionsetup{width=\textwidth}\caption{Graphical illustration of the local refinement of the initial mesh from the three EOSs with smallest values of $e_i$. The image represents a simplified model with $2$ polytropic parameters $X=[1,8]_8$ and $Y=[1,9]_9$.}\label{fig6}
\end{figure}

The algorithm, which will be called the piecewise polytropic meshing and refinement method for the inverse problem, proceeds as follows:
\begin{enumerate}
	\item calculate the fundamental $wI$ mode ($\nu$, $\tau$) and the mass ($M$) for every EOS in the $4$-volume of polytropic parameters. We calculate $20$ configurations for each EOS in the same fixed central pressure range, namely $\log_{10}p_0=[33.9,36.61]_{20}$.
	\item fit with piecewise linear interpolation the curve $\nu_i(M_i)$ for all the EOSs in the $4$-space of polytropic parameters. The interpolation is necessary to calculate $\nu_i(M_{\text{exp}})$.
	\item compare each curve $\nu_i(M_i)$ with the input data $\nu_{\text{exp}}(M_{\text{exp}})$ by calculating
	\begin{myequation}
		e_i=\max\left\{\frac{\abs{\nu_i(M_{\text{exp}})-\nu_{\text{exp}}(M_{\text{exp}})}}{\nu_{\text{exp}}(M_{\text{exp}})}\right\}.
	\end{myequation}
	
	We only calculate $e_i$ for those EOSs that fulfill the condition $\max (M_i)\geq\max (M_{\text{exp}})$. The smaller $e_i$ is, the more similar $\nu_i(M_i)$ and $\nu_{\text{exp}}(M_{\text{exp}})$ are.
	\item sort the EOSs in increasing order of $e_i$ and check the value of $\min_i(e_i)$.
	\item if $\min_i(e_i)<\text{tol}$, finish the algorithm. In other case, define a new $4$-volume of polytropic parameters that contains, for example, the first $3$ EOSs with smallest $e_i$, and repeat from step $1$. This new $4$-volume of polytropic parameters is a local refinement of the initial mesh. A graphical illustration of the local refinement is shown in Fig. \ref{fig6}.
\end{enumerate}

\onecolumngrid

\twocolumngrid

\begin{figure}[H]
\centering
\includegraphics[width=\linewidth]{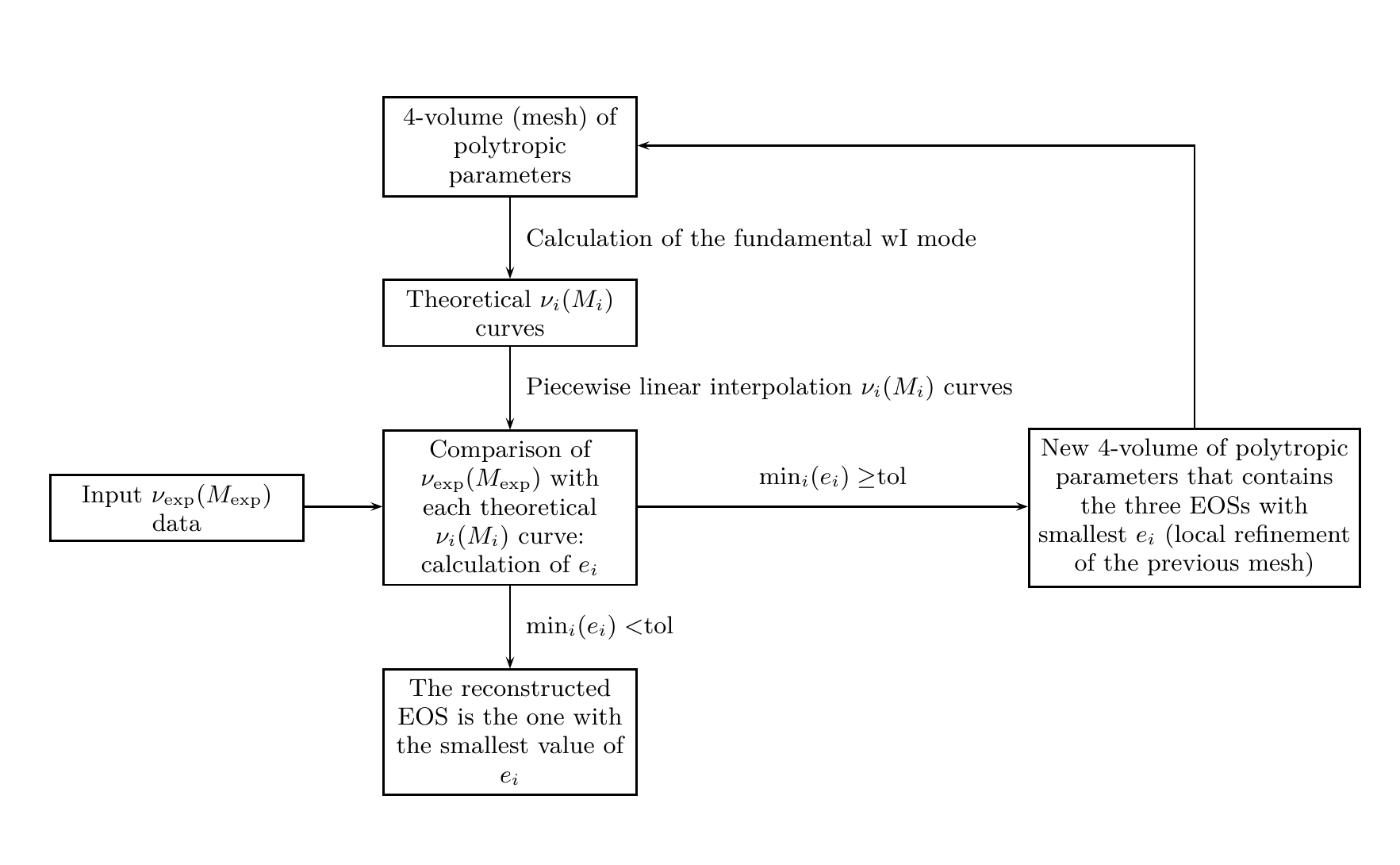}\captionsetup{width=\textwidth}\caption{Scheme of the piecewise polytropic meshing and refinement method for the inverse problem.}\label{fig7}
\end{figure}

A scheme of the entire meshing and refinement method is shown in FIG. \ref{fig7}.

\onecolumngrid

\twocolumngrid

The main difficulty of the meshing and refinement method is to calculate so many quasinormal modes ($20$ for each EOS, which means a total of $20\times 27440=548800$ quasinormal modes in the initial mesh). To obtain an only quasinormal mode for a given EOS, one has to find a complex frequency $\omega$ that verifies the junction condition \eqref{junction} when solving the Regge-Wheeler equation inside and outside the neutron star. This gives rise to many convergence problems, but we got the Müller method working quite correctly.

Once the algorithm has finished, we will have a reconstructed EOS. Then, we will be able to calculate other macroscopic parameters (such as the moment of inertia, the quadrupole moment, the tidal Love parameter,\dots) in order to make predictions that can be compared with other experimental data.

Since today we still do not have the necessary experimental data, in order to test the algorithm we will suppose that the measured macroscopic parameters correspond to, for example, the ones calculated for GNH3 EOS ($M_{\text{exp}}=M_{\text{GNH3}}$ and $\nu_{\text{exp}}=\nu_{\text{GNH3}}$). We will consider $5$ of the $20$ GNH3 configurations shown in Fig. \ref{fig2}. Since GNH3 is a known EOS, we will be able to directly compare the reconstructed EOS with the original one and also to compare them in macroscopic parameters.

In Sec. \ref{resultadosII} we plot some macroscopic parameters that we calculated for the $27440$ EOSs in the initial mesh of polytropic parameters. The numerical results of the meshing and refinement method are described in Sec. \ref{problema_inverso_resultados}.

\section{Numerical results for the 27440 equations of state generated}\label{resultadosII}

In this section, we will show relations between some of the macroscopic parameters of the $27440$ EOSs generated in the $4$-volume of polytropic parameters defined by Eq. \eqref{parametrosI}.

\begin{figure}[H]
	\centering
	\includegraphics[width=\linewidth]{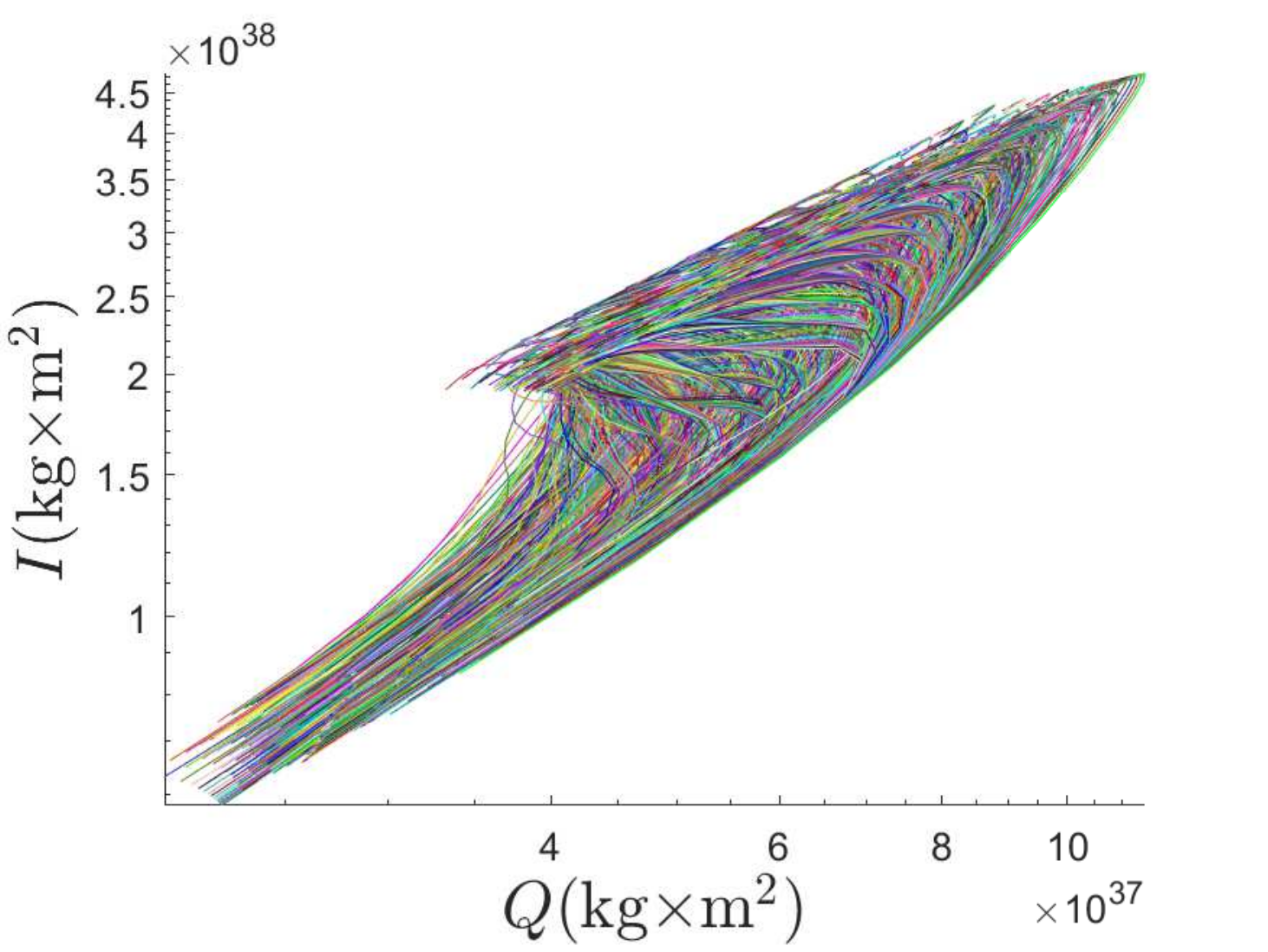}\captionsetup{width=0.5\textwidth}\caption{Moment of inertia vs quadrupole moment for the $27440$ EOSs generated (colors randomly chosen).}\label{fig8}
\end{figure}

In Fig. \ref{fig8} we plot the relation between the moment of inertia and the quadrupole moment for the $27440$ EOSs generated. We observe that each EOS has its own behavior.

\begin{figure}[H]
	\centering
	\includegraphics[width=\linewidth]{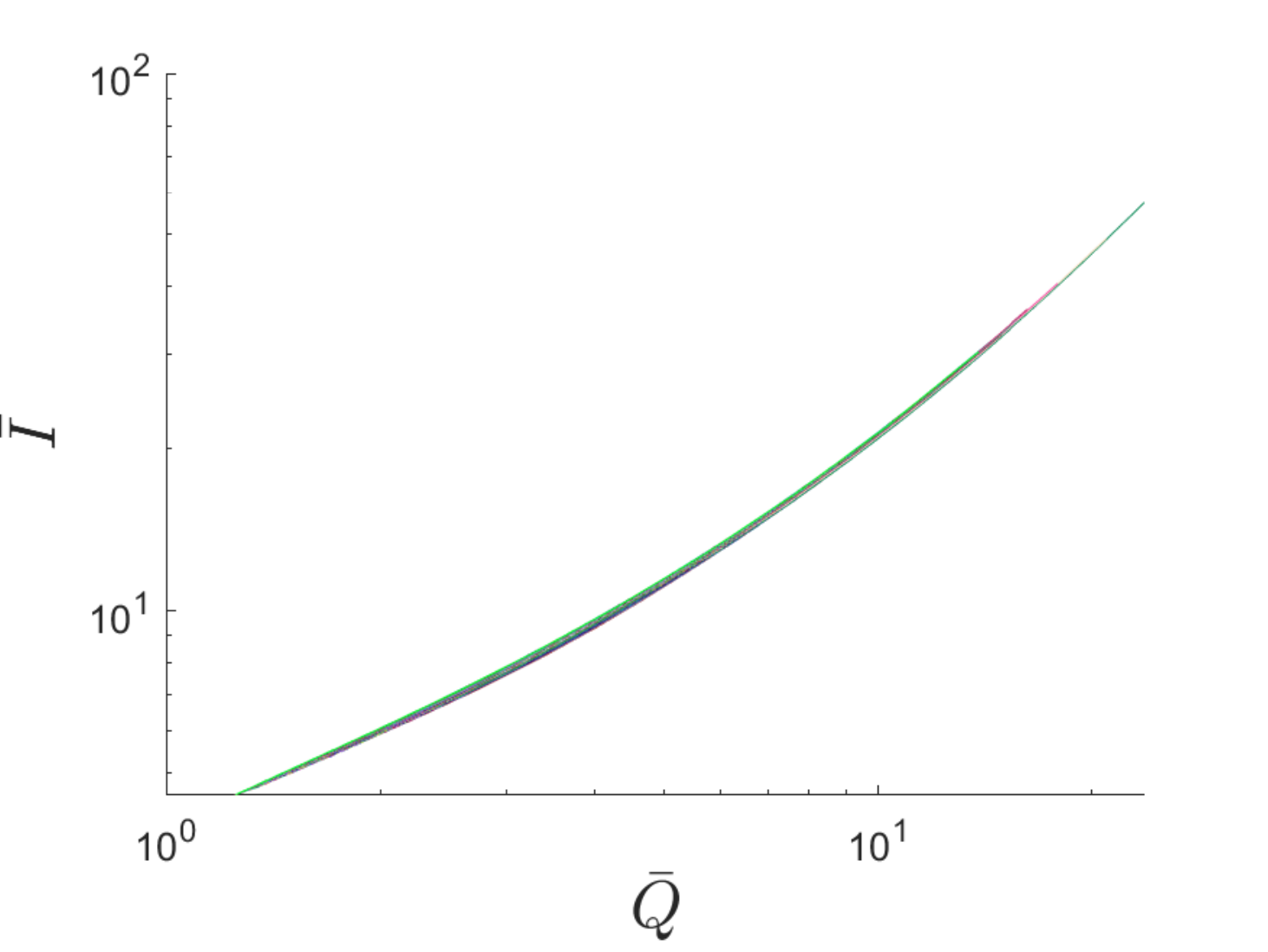}\captionsetup{width=0.5\textwidth}\caption{I-Q relation for the $27440$ EOSs generated (colors randomly chosen).}\label{fig9}
\end{figure}

\begin{figure}[H]
	\centering
	\includegraphics[width=\linewidth]{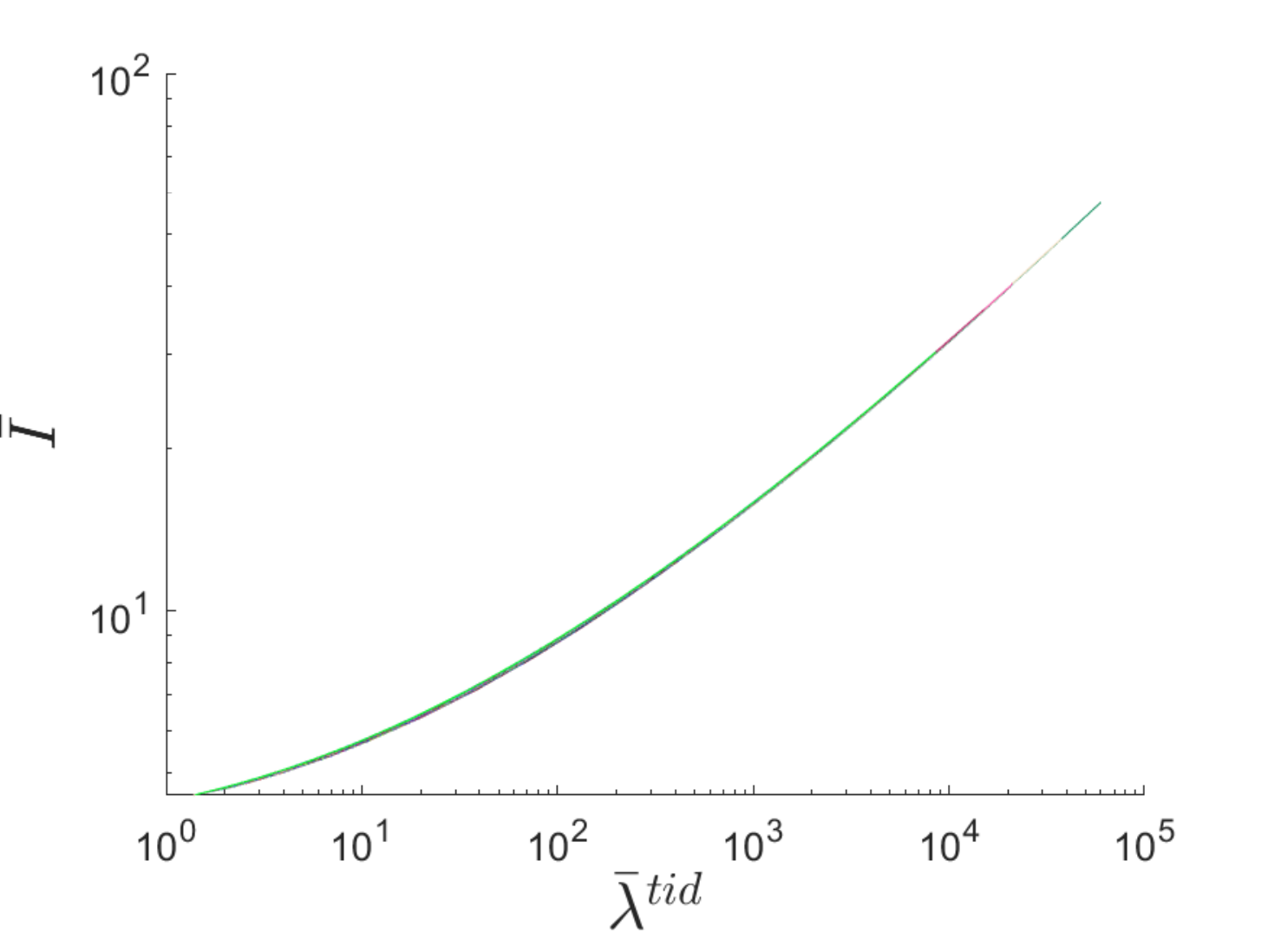}\captionsetup{width=0.5\textwidth}\caption{I-Love relation for the $27440$ EOSs generated (colors randomly chosen).}\label{fig10}
\end{figure}

However, a universal relation between these two parameters is found if they are re-scaled as in Eq. \eqref{IQparameters}. Also considering the tidal Love parameter leads to the so-called I-Love-Q relations. In Figs. \ref{fig9} and \ref{fig10} we plot the I-Q and the I-Love relations, respectively.

The two previous relations are EOS independent: the $27440$ EOSs fall in the same curve. The fact of having so many different EOSs falling in the same curve only increases the universality of the I-Love-Q relations.

\begin{figure}[H]
	\centering
	\includegraphics[width=\linewidth]{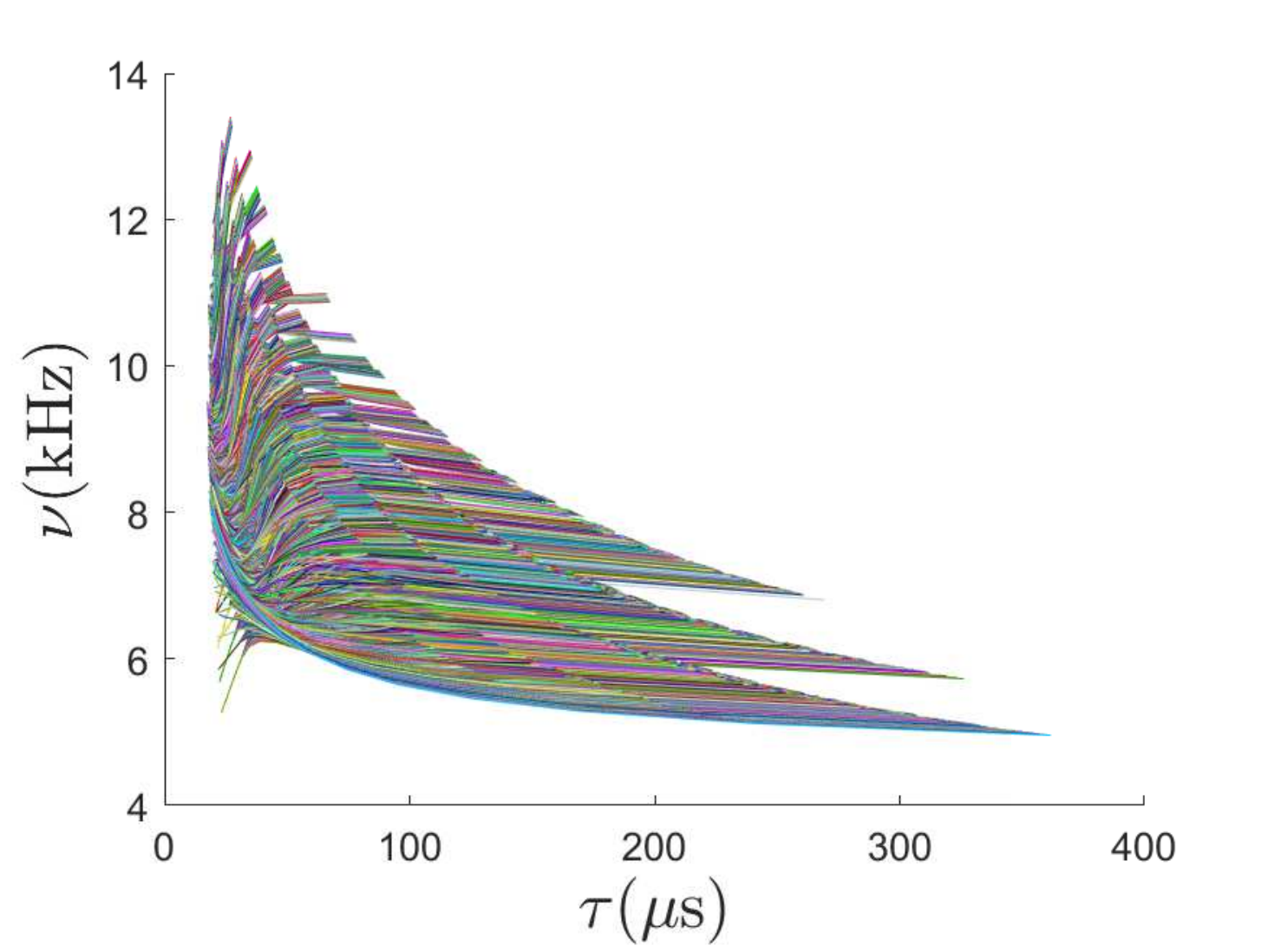}\captionsetup{width=0.5\textwidth}\caption{Frequency of the fundamental $wI$ mode vs its damping time for the $27440$ EOSs generated (colors randomly chosen).}\label{fig11}
\end{figure}

As for the quasinormal modes, in Fig. \ref{fig11} we plot the relation between the frequency of the fundamental $wI$ mode and its damping time.

As seen in Fig. \ref{fig11}, the relation between the frequency of the fundamental $wI$ mode (given by $2\pi\nu=\Re\omega$) and its damping time (given by $\tau^{-1}=\Im\omega$) clearly depends on the EOS. However, one may find universal relations between $\Im\omega$ and $\Re\omega$ if $\omega$ is properly re-scaled. 

\begin{figure}[H]
	\centering
	\includegraphics[width=\linewidth]{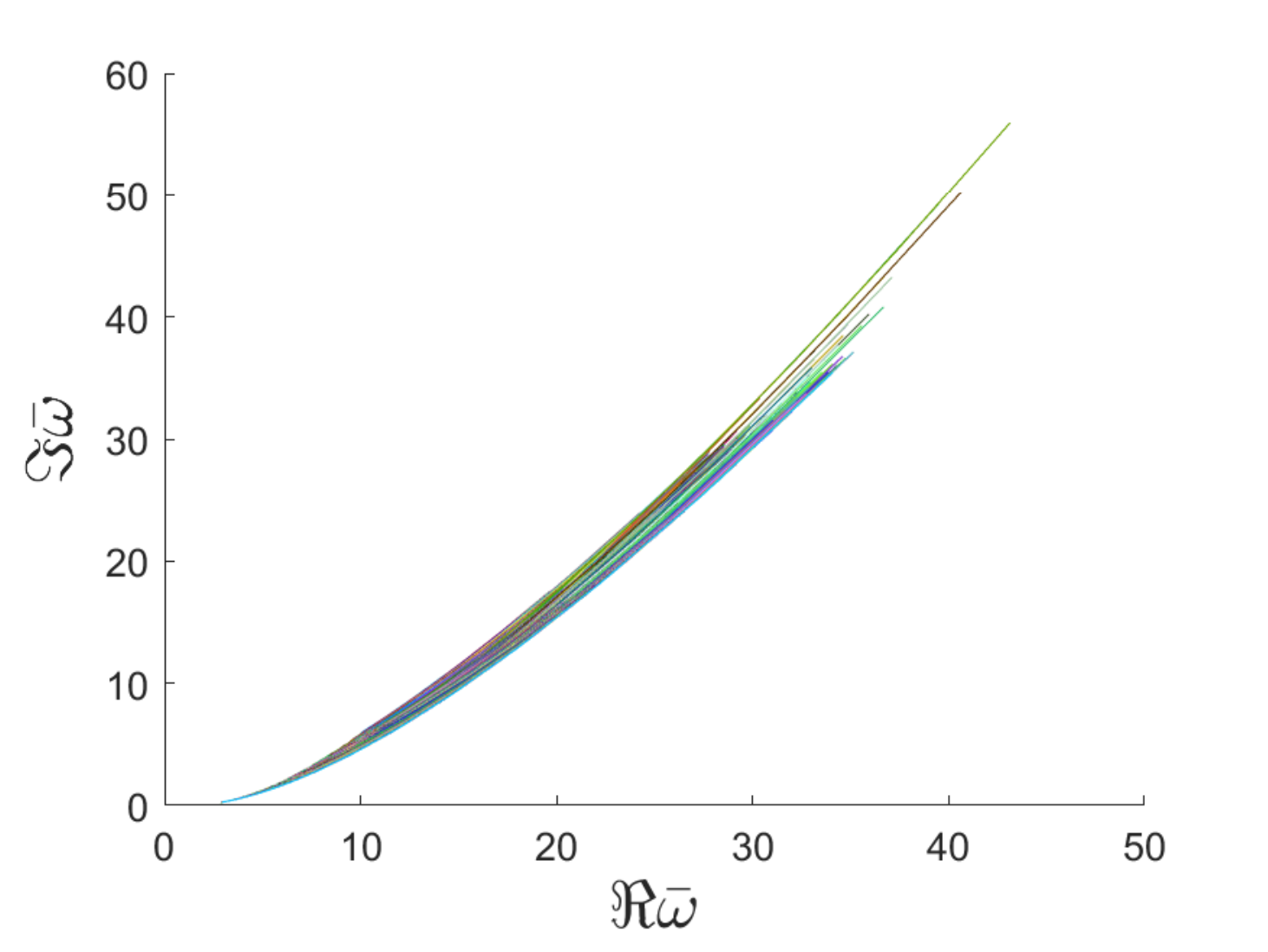}\captionsetup{width=0.5\textwidth}\caption{$\Im\bar{\omega}$ vs $\Re\bar{\omega}$ for the $27440$ EOSs generated (colors randomly chosen). The scaling with the square root of the central pressure is quite independent of the EOS.}\label{fig12}
\end{figure}

In Ref. \cite{blazquez2013phenomenological} it was shown that re-scaling $\omega$ with the square root of the central pressure,
\begin{myequation}
	\bar{\omega}=\frac{c}{\sqrt{Gp_0}}\omega,
\end{myequation}
leads to an approximately-universal relation. This is shown in Fig. \ref{fig12}.

\begin{figure}[H]
	\centering
	\includegraphics[width=\linewidth]{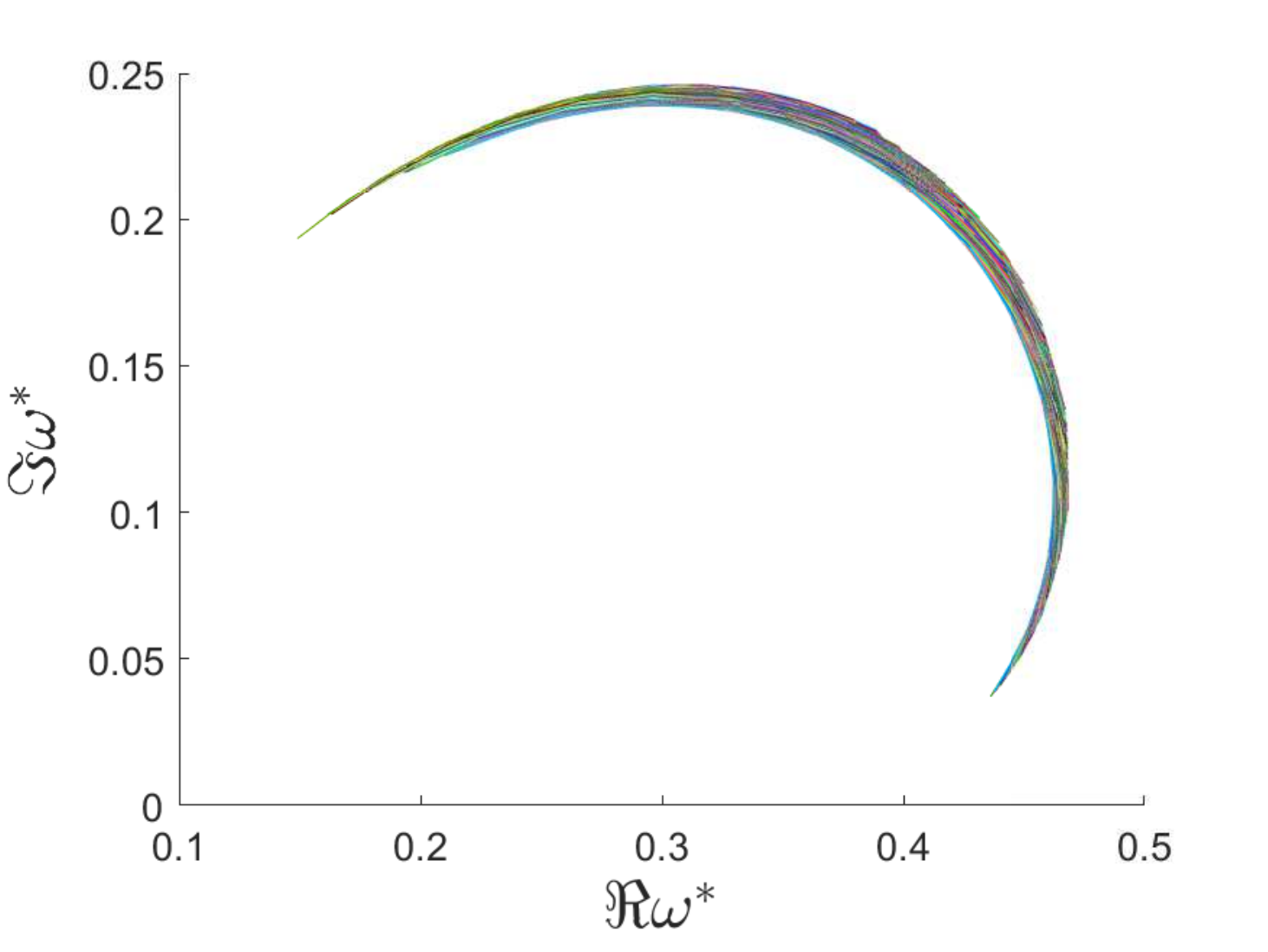}\captionsetup{width=0.5\textwidth}\caption{$\Im\omega^*$ vs $\Re\omega^*$ for the $27440$ EOSs generated (colors randomly chosen). The scaling with the mass is quite independent of the EOS.}\label{fig13}
\end{figure}

We can also find universal relations if $\omega$ is re-scaled with the mass,
\begin{myequation}
	\omega^*=\frac{GM}{c^3}\omega.
\end{myequation}
This universal relation is shown in Fig. \ref{fig13}.

\begin{figure}[H]
	\centering
	\includegraphics[width=\linewidth]{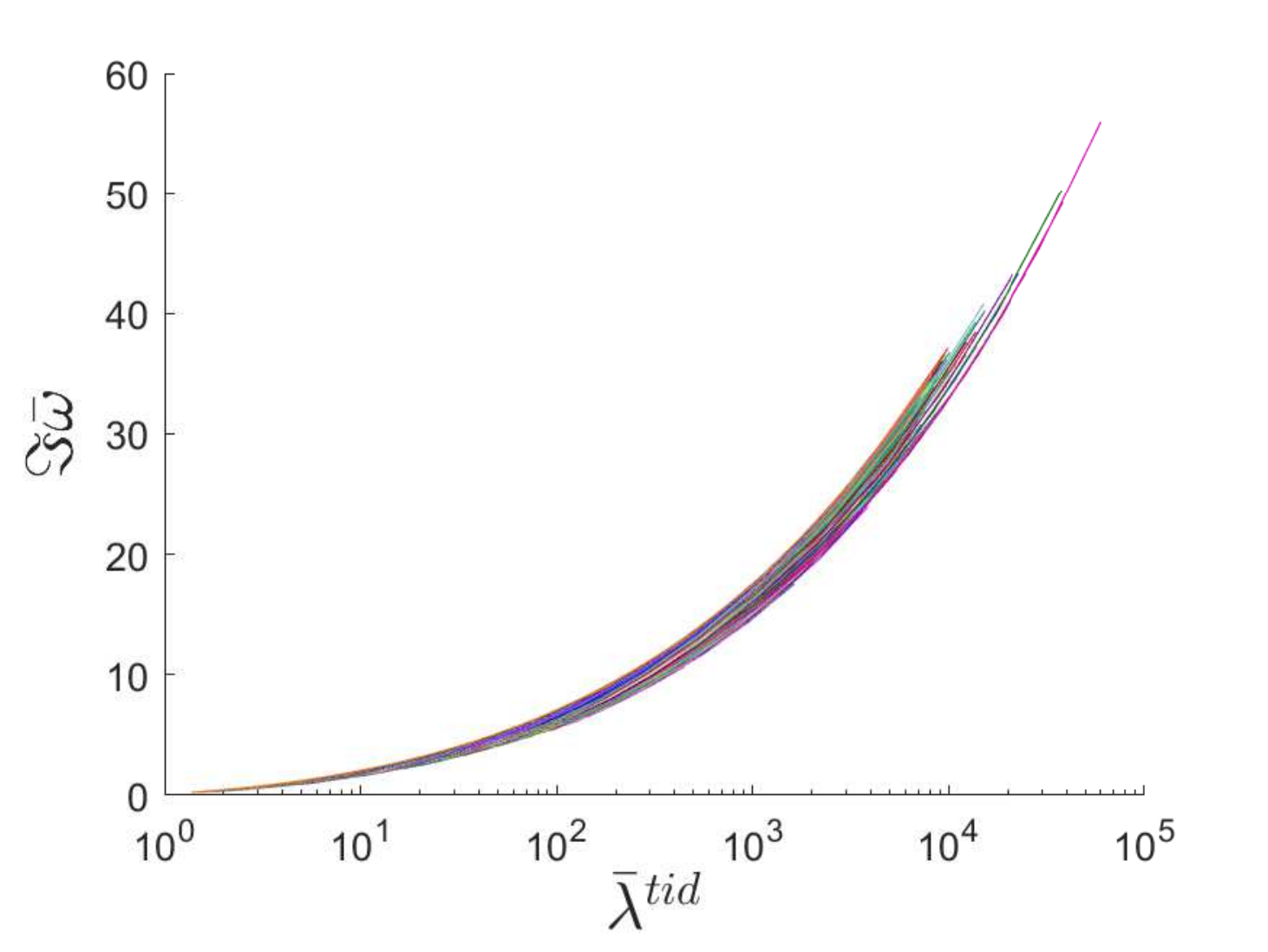}\captionsetup{width=0.5\textwidth}\caption{$\Im\bar{\omega}$ vs $\bar{\lambda}^{tid}$ for the $27440$ EOSs generated (colors randomly chosen).}\label{fig14}
\end{figure}

\begin{figure}[H]
	\centering
	\includegraphics[width=\linewidth]{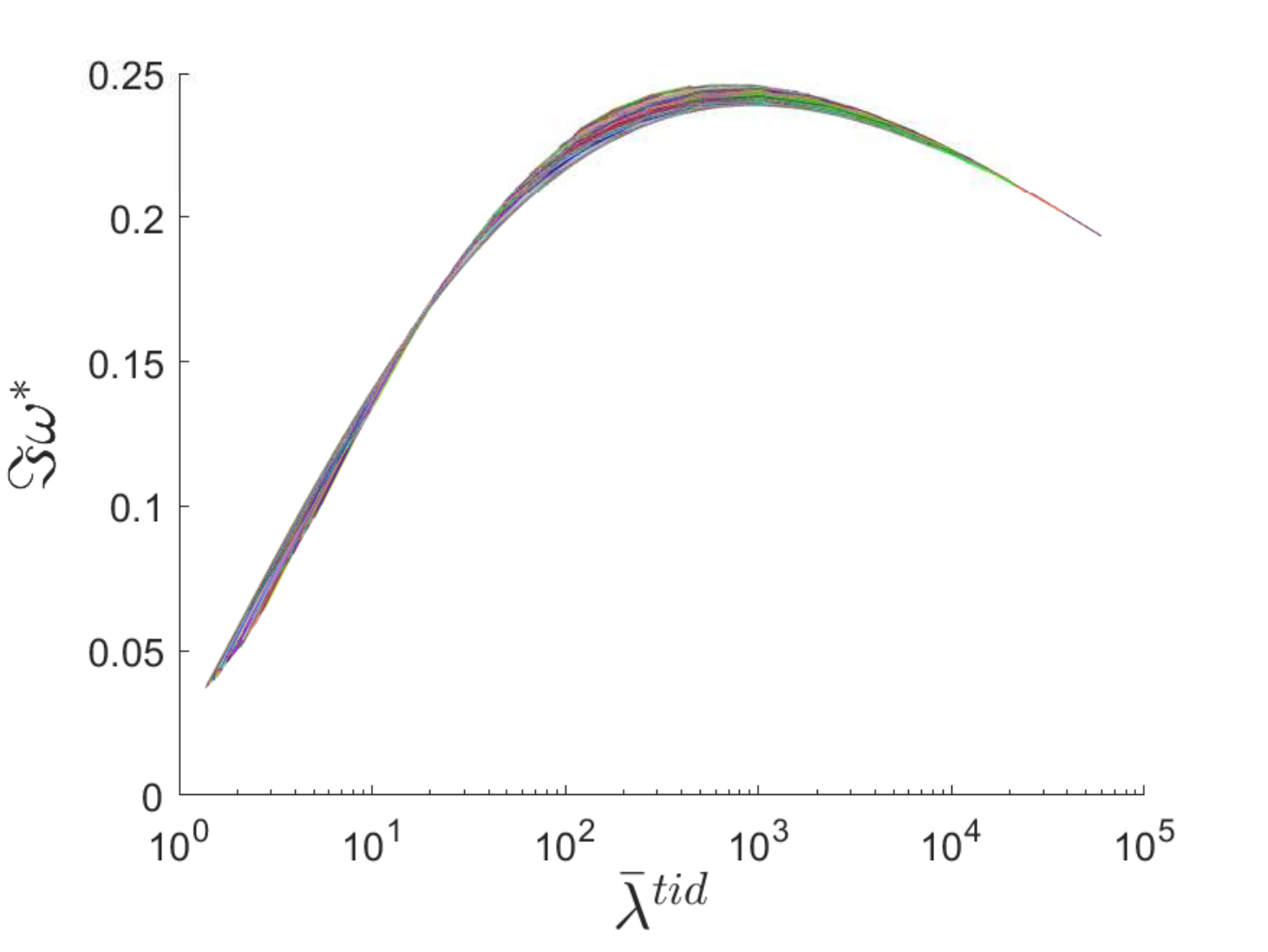}\captionsetup{width=0.5\textwidth}\caption{$\Im\omega^*$ vs $\bar{\lambda}^{tid}$ for the $27440$ EOSs generated (colors randomly chosen).}\label{fig15}
\end{figure}

One may also find other interesting universal relations between $\bar{\omega}$ (or $\omega^*$) and $\bar{\lambda}^{tid}$. In Figs. \ref{fig14} and \ref{fig15} we plot the relation between $\Im\bar{\omega}$ and $\bar{\lambda}^{tid}$, and between $\Im\omega^*$ and $\bar{\lambda}^{tid}$, respectively.

\begin{figure}[H]
\centering
\includegraphics[width=\linewidth]{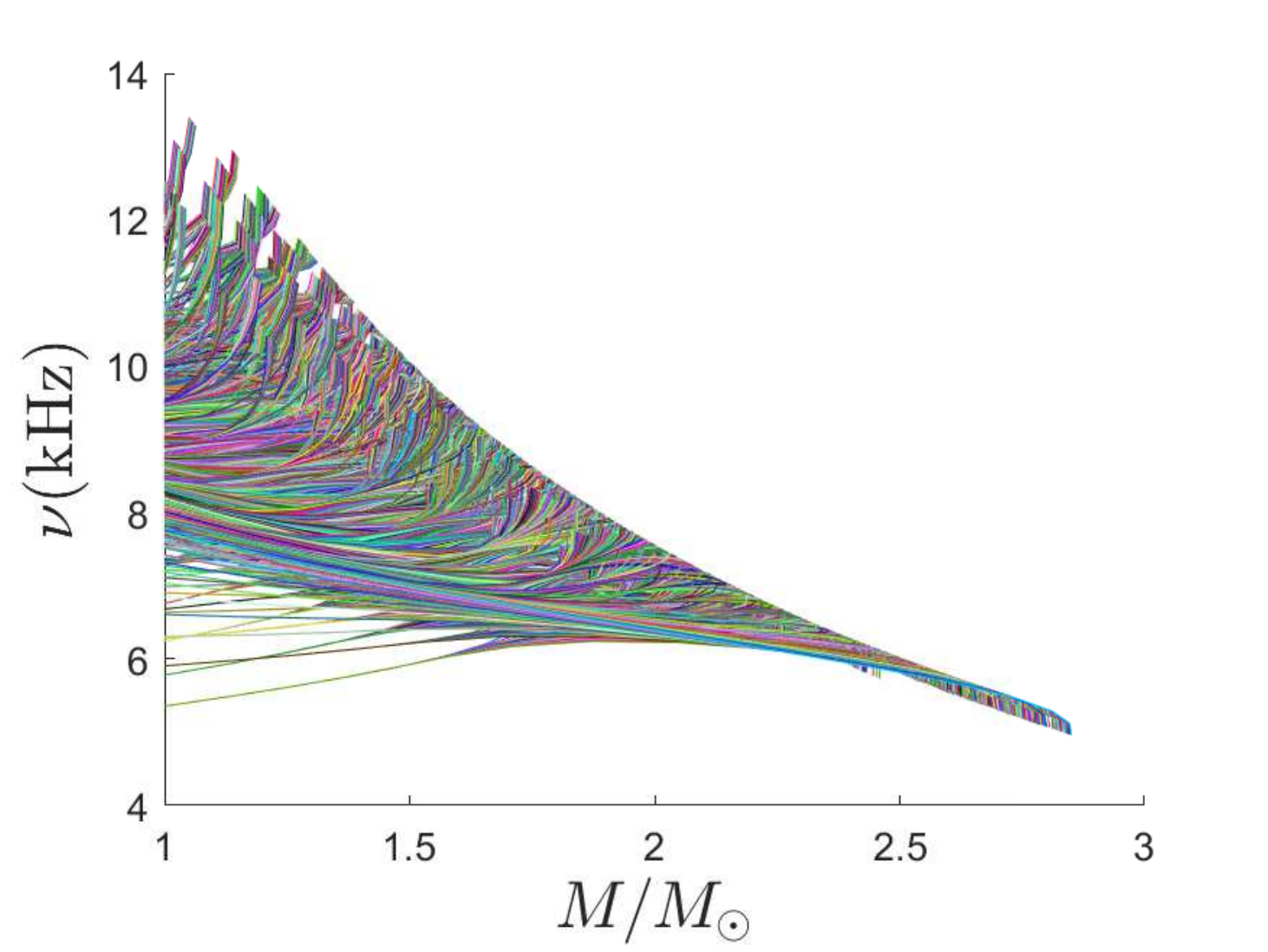}\captionsetup{width=0.5\textwidth}\caption{Frequency of the fundamental $wI$ mode vs mass for the $27440$ EOSs generated (colors randomly chosen).}\label{fig16}
\end{figure}

Finally, in Fig. \ref{fig16} we plot the curves $\nu_i(M_i)$, $i=1,\dots,27440$. As explained in Sec. \ref{problema_inverso}, these curves are essential to develop our approach to the inverse problem.

\section{Testing the meshing and refinement method}\label{problema_inverso_resultados}

In this section, we will test the piecewise polytropic meshing and refinement method explained in Sec. \ref{problema_inverso}. We will consider some GNH3 EOS stellar configurations as our input data, i.e., $M_{\text{exp}}=M_{\text{GNH3}}$ and $\nu_{\text{exp}}=\nu_{\text{GNH3}}$. To carry out the test we will consider a tolerance $\text{tol}=0.005$.

\vspace{0.07cm}\begin{center} \textbf{First iteration of the method}\end{center}\vspace{0.07cm}

We proceed as explained in Sec. \ref{problema_inverso} (steps 1. to 4.). The polytropic parameters of the first $3$ EOSs with the smallest values of $e_i$ in the mesh of EOSs given by Eq. \eqref{parametrosI} are listed in Table \ref{tabla1}.

\begin{table}[H]
	\begin{center}
		\begin{tabular}{| c | c | c | c | c |}\hline
			$e_i$ & $\log_{10} p_1$ & $\Gamma_1$ & $\Gamma_2$ & $\Gamma_3$\\\hline
			0.0078506 & 34.644 & 2.6923 & 2.2615 & 1.9538\\
			0.010446 & 34.644 & 2.5538 & 2.2615 & 2.1077\\
			0.011615 & 34.644 & 2.6923 & 2.2615 & 2.1077\\\hline
		\end{tabular}
	\end{center}
	\captionsetup{width=0.5\textwidth}\caption{Polytropic parameters of the first 3 EOSs with the smallest values of $e_i$ in the $4$-volume defined by Eq. \eqref{parametrosI}.}\label{tabla1}
\end{table}

Since $\min_i(e_i)\geq 0.005$, we proceed with the refinement of the initial mesh (step 5. of our method). Hence, we define a local refinement of the initial $4$-volume of piecewise polytropic parameters, which was given by Eq. \eqref{parametrosI}, that contains the EOSs listed in Table \ref{tabla1}. This local refinement is given by
\begin{myequation}
	\begin{array}{ll}
		\log_{10}p_1=34.644,\\\\
		\Gamma_1=[2.5538,2.6923],\\\\
		\Gamma_2=2.2615,\\\\
		\Gamma_3=[1.9538,2.1077].
	\end{array}
\end{myequation}
Note that we have one single element in $\log_{10}p_1$ and $\Gamma_2$. In order to avoid this, we will consider the adyacent values for all the parameters, i.e.,
\begin{myequation}\label{parametrosII}
	\begin{array}{ll}
		\log_{10}p_1=[34.589,34.7]_{8},\\\\
		\Gamma_1=[2.4154,2.8308]_{10},\\\\
		\Gamma_2=[2.1077,2.4154]_{10},\\\\
		\Gamma_3=[1.8,2.2615]_{10}.
	\end{array}
\end{myequation}
A graphical illustration of the local refinement of the initial mesh was shown in Fig. \ref{fig6}. The new $4$-volume will contain a total of $8\times 10^3=8000$ EOSs. 

\vspace{0.07cm}\begin{center} \textbf{Second iteration of the method (first refinement)}\end{center}\vspace{0.07cm}

Again, we proceed as explained in the algorithm in Sec. \ref{problema_inverso} (steps 1. to 4.). The polytropic parameters of the first $4$ EOSs with the smallest values of $e_i$ in the mesh of EOSs given by Eq. \eqref{parametrosII} are listed in Table \ref{tabla2}. We list 4 EOSs instead of 3 because the first two ones have the same value of $e_i$.

\begin{table}[H]
	\begin{center}
		\begin{tabular}{| c | c | c | c | c |}\hline
			$e_i$ & $\log_{10} p_1$ & $\Gamma_1$ & $\Gamma_2$ & $\Gamma_3$\\\hline
			0.0050477 & 34.637 & 2.5539 & 2.2786 & 2.1077\\
			0.0050477 & 34.637 & 2.5539 & 2.2786 & 2.1589\\
			0.0058107 & 34.652 & 2.6923 & 2.1761 & 2.2615\\
			0.0065108 & 34.637 & 2.5539 & 2.2786 & 2.2102\\\hline
		\end{tabular}
	\end{center}
	\captionsetup{width=0.5\textwidth}\caption{Polytropic parameters of the first 4 EOSs with the smallest values of $e_i$ in the $4$-volume defined by Eq. \eqref{parametrosII}.}\label{tabla2}
\end{table}

Since $\min_i(e_i)\geq 0.005$, we proceed with the refinement of the initial mesh (step 5. of our method). We now define a local refinement of the previous $4$-volume of polytropic parameters, which was given by Eq. \eqref{parametrosII}, from the EOSs listed in Table \ref{tabla2}. This new local refinement is given by

\begin{myequation}\label{parametrosIII}
	\begin{array}{ll}
		\log_{10}p_1=[34.621,34.668]_{8},\\\\
		\Gamma_1=[2.5077,2.7385]_{10},\\\\
		\Gamma_2=[2.1419,2.3128]_{10},\\\\
		\Gamma_3=[2.0564,2.3128]_{10}.
	\end{array}
\end{myequation}
In the third iteration we will have a total of $8\times 10^3=8000$ EOSs. 

\vspace{0.07cm}\begin{center} \textbf{Third iteration of the method (second refinement)}\end{center}\vspace{0.07cm}

Again, we proceed as explained in the algorithm in Sec. \ref{problema_inverso} (steps 1. to 4.). The polytropic parameters of the first $4$ EOSs with the smallest values of $e_i$ in the mesh of EOSs given by Eq. \eqref{parametrosIII} are listed in Table \ref{tabla3}. As in the second iteration, we list 4 EOSs instead of 3 because the first two ones have the same value of $e_i$.

\begin{table}[H]
	\begin{center}
		\begin{tabular}{| c | c | c | c | c |}\hline
			$e_i$ & $\log_{10} p_1$ & $\Gamma_1$ & $\Gamma_2$ & $\Gamma_3$\\\hline
			0.0035480 & 34.648 & 2.6616 & 2.1989 & 2.2843\\
			0.0035480 & 34.648 & 2.6616 & 2.1989 & 2.3128\\
			0.0037085 & 34.648 & 2.6359 & 2.1989 & 2.2843\\
			0.0037134 & 34.648 & 2.6359 & 2.1989 & 2.3128\\\hline
		\end{tabular}
	\end{center}
	\captionsetup{width=0.5\textwidth}\caption{Polytropic parameters of the first 4 EOSs with the smallest values of $e_i$ in the $4$-volume defined by Eq. \eqref{parametrosIII}.}\label{tabla3}
\end{table}

Since $\min_i(e_i)<\text{tol}$, we stop the algorithm. From now on, the first EOS listed in Table \ref{tabla3} will be denoted as the reconstructed GNH3 EOS.

\vspace{0.07cm}\begin{center} \textbf{Comparison between the original and the reconstructed GNH3 equations of state}\end{center}\vspace{0.07cm}

Here we will distinguish between three different GNH3 EOS:
\begin{itemize}
	\item The original GNH3 EOS \cite{glendenning1984neutron}.
	\item The reconstructed GNH3 EOS. This is the one our algorithm reconstructed, whose polytropic parameters are listed in the first row of Table \ref{tabla3}.
	\item The polytropic GNH3 EOS. This one is the polytropic fit of GNH3 EOS, whose polytropic parameters can be found in Table III of Ref. \cite{read2009constraints}.
\end{itemize}

\begin{figure}[H]
	\centering
	\includegraphics[width=\linewidth]{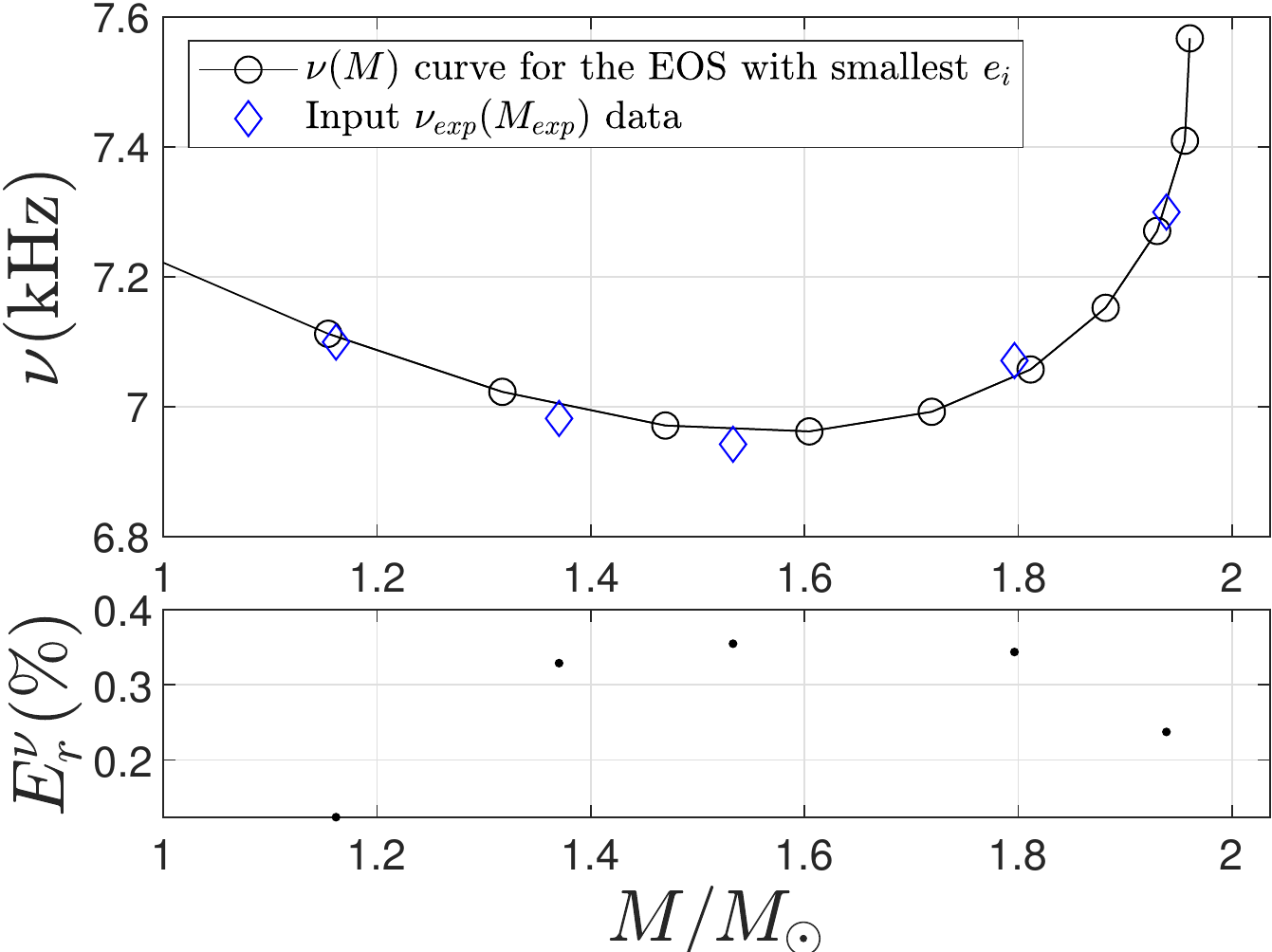}
	\captionsetup{width=0.5\textwidth}\caption{
Top panel: frequency of the fundamental $wI$ mode vs mass for the input data (blue diamonds) and the reconstructed EOS (black circles). Bottom panel: relative difference between the input data and the reconstructed EOS.}\label{fig17}
\end{figure}

Fig. \ref{fig17} shows the frequency of the fundamental $wI$ mode vs the mass for the original GNH3 EOS (blue diamonds) and for the reconstructed GNH3 EOS (black circles), together with the relative difference.

Since the input data correspond to $5$ GNH3 configurations, we can compare the reconstructed GNH3 EOS with the original one (this would not be possible if we had used real experimental data). There are two different ways to compare them: 
\begin{enumerate}
\item Directly comparing the EOSs. The easiest way to do it is by comparing the polytropic parameters of the reconstructed EOS with those of the polytropic one. The polytropic parameters of both EOSs are listed in Table \ref{tabla4}.
	
\begin{table}[H]
\begin{center}
\begin{tabular}{| c | c | c | c | c |}\hline
EOS & $\log_{10} p_1$ & $\Gamma_1$ & $\Gamma_2$ & $\Gamma_3$\\\hline
Reconstructed GNH3 & 34.648 & 2.6616 & 2.1989 & 2.2843\\
Polytropic GNH3 & 34.648 & 2.664 & 2.194 & 2.304\\\hline
\end{tabular}
\end{center}
\captionsetup{width=0.5\textwidth}\caption{Comparison between the polytropic parameters of the reconstructed GNH3 EOS and the polytropic one.}\label{tabla4}
\end{table}
	
The polytropic parameters of the reconstructed GNH3 EOS are very similar to those of the polytropic one, as expected.
	
\item Comparing both EOSs in macroscopic parameters by numerically solving the equations obtained in Sec. \ref{overview} and in the Appendices \ref{appendixA} and \ref{appendixB}. In Table \ref{tabla5} we compare the reconstructed GNH3 EOS with the original one by calculating $20$ stellar configurations with the same central energy densities for both EOSs. We also compare the original GNH3 EOS with the polytropic one, in order to check how good our reconstruction is.

The reconstructed GNH3 EOS is very similar to the original one in macroscopic parameters. Comparing the results of both columns of Table \ref{tabla5}, we conclude that our polytropic reconstruction of GNH3 EOS is almost as good as the polytropic fit of Ref. \cite{read2009constraints}.
	
\begin{table}[H]
\begin{center}
\begin{tabular}{| c | c | c |}\hline
\multirow{ 2}{*}{Parameter} & \multicolumn{2}{c|}{Maximum relative difference ($\%$)}\\\cline{2-3}
& a) Original-reconstructed & b) Original-polytropic\\\hline
$p_0$ & $2.4249$ & $2.2727$\\
$R$ & $0.62698$ & $0.63248$\\
$M$ & $0.91586$ & $0.86603$\\
$I$ & $1.3854$ & $1.0864$\\
$\bar{I}$ & $1.7485$ & $1.6217$\\
$Q$ & $1.9447$ & $1.9695$\\
$\bar{Q}$ & $2.1123$ & $1.7168$\\
$\bar{\lambda}^{tid}$ & $6.316$ & $5.8184$\\
$\nu$ & $0.44696$ & $0.41161$\\
$\tau$ & $1.2084$ & $1.0827$\\
$\Re \bar{\omega}$ & $1.2664$ & $1.1411$\\
$\Im \bar{\omega}$ & $2.3122$ & $2.0984$\\
\hline
\end{tabular}
\end{center}
\captionsetup{width=0.5\textwidth}\caption{a) Maximum relative difference in macroscopic parameters between the original GNH3 EOS and the reconstructed one out of $20$ configurations calculated. b) Maximum relative difference in macroscopic parameters between the original GNH3 EOS and the polytropic one out of $20$ configurations calculated.}\label{tabla5}
\end{table}

\end{enumerate}

\begin{figure}
\includegraphics[width=\linewidth]{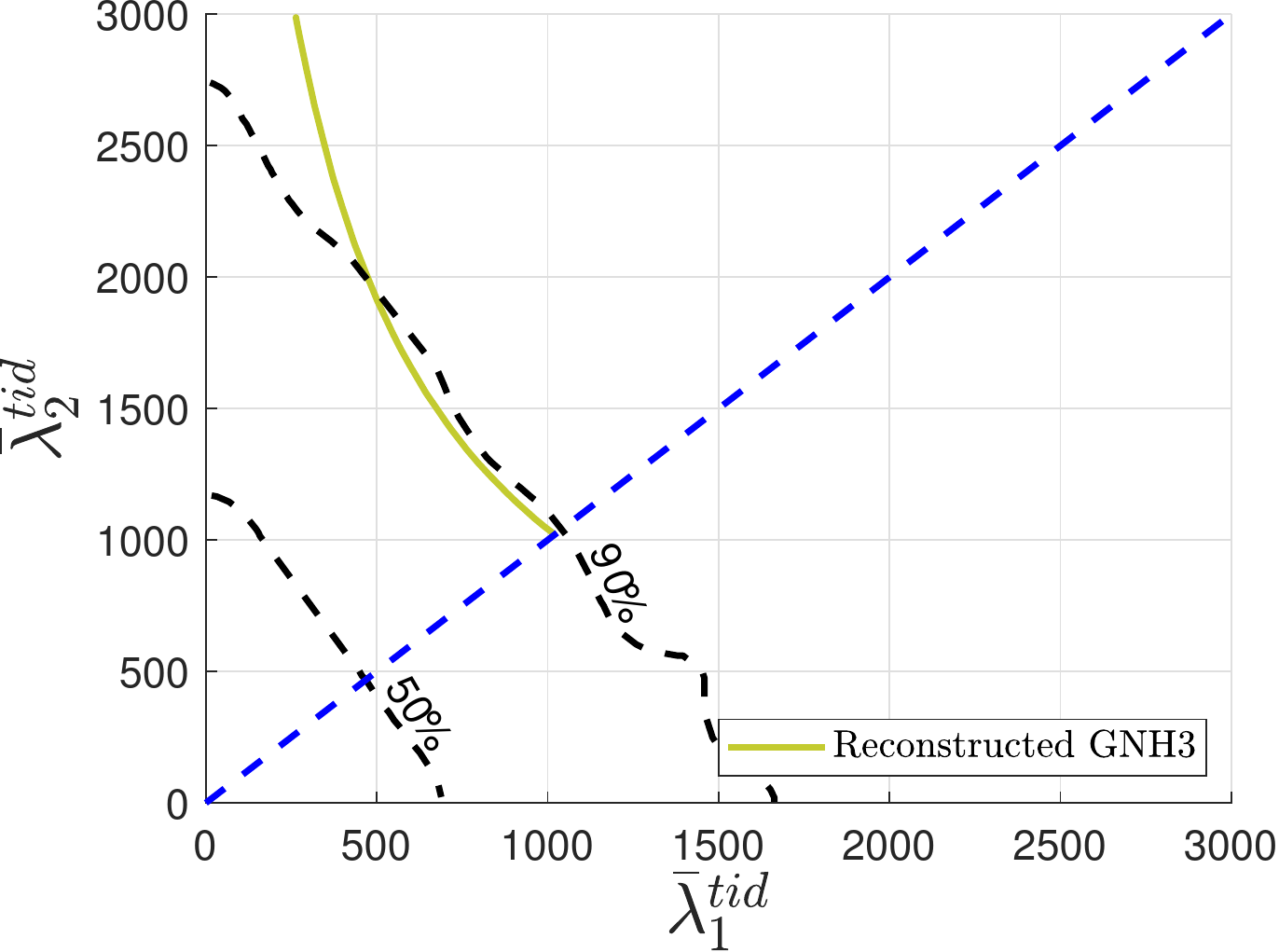}\captionsetup{width=0.5\textwidth,justification=justified}\caption{Predictions for tidal deformability given by the reconstructed GNH3 EOS, under the assumption that both components are neutron stars. Contours enclosing $90\%$ and $50\%$ of the probability density are shown as dashed lines (both curves taken from Ref. \cite{abbott2017gw170817}).}\label{fig18}
\end{figure}

Now that the neutron star EOS has been determined, we can also calculate the predictions for the tidal deformability given by the reconstructed EOS, as we did in Fig. \ref{fig3} for the realistic EOSs considered in this paper. This analysis would be more interesting if the input data were real experimental data and not GNH3 configurations, since then we would check if the reconstructed EOS would predict $\bar{\lambda}^{tid}$ values inside the $90\%$ confidence contour or not. Since the input data are GNH3 configurations, we expect the results to be very similar to the ones obtained for GNH3 EOS. The resulting curve is shown in Fig. \ref{fig18}.

We conclude that our reconstruction of GNH3 EOS with the meshing and refinement method is almost as good as the piecewise polytropic fit. This means that, starting from only $5$ input $(M,\nu)$ points, we have been able to reconstruct the neutron star EOS in a very good approximation.

We also applied the meshing and refinement method for several EOSs (APR4, SLy, ALF4,\dots), and we found that GNH3 EOS is, in fact, one of the most conflictive EOS, i.e., the number of iterations needed for any of these EOSs is smaller the ones needed for GNH3.

\subsection{Testing the meshing and refinement method with experimental error}

Consider the same $5$ GNH3 configurations we used as input data for the meshing and refinement method. If the input data was experimental data, it would have an associated experimental error. In order to make an estimation of how this experimental error would affect the final results, now we will randomly modify the input data in a uncertainty interval, i.e.,
\begin{myequation}\label{random_input}
X^i_{\text{exp}}\to X^i_{\text{exp}}+\text{rand}\left[-\Delta X^i_{\text{exp}},\Delta X^i_{\text{exp}}\right],\ i=1,\dots,5,
\end{myequation}
with 
\begin{myequation}
\Delta X_{\text{exp}}^i=\epsilon_X X_{\text{exp}}^i,\ i=1,\dots,5,
\end{myequation}
where $X$ is either $M$ or $\nu$, and rand$\left[A,B\right]$ represents the standard uniform distribution in the interval $[A,B]$. We will consider the same error $\epsilon_M=\epsilon_{\nu}=0.01$ for both the mass and the frequency of the fundamental $wI$ mode.

The objective of this analysis is to find out how an experimental error, i.e., $\Delta M_{\text{exp}}$ and $\Delta \nu_{\text{exp}}$, would propagate to the polytropic parameters of the reconstructed equation of state.

We will carry out several realizations of the first iteration of the meshing and refinement method with different inputs. We here present three typical realizations to show the characteristics of the results.

\vspace{0.07cm}\begin{center} \textbf{First realization of the test}\end{center}\vspace{0.07cm}

The results of the first realization of the test are shown in Table \ref{tabla6}.

\begin{table}[H]
	\begin{center}
		\begin{tabular}{| c | c | c | c | c |}\hline
			$e_i$ & $\log_{10} p_1$ & $\Gamma_1$ & $\Gamma_2$ & $\Gamma_3$\\\hline
			0.010565 & 34.644 & 2.5538 & 2.2615 & 2.1077\\
			0.011084 & 34.644 & 2.5538 & 2.2615 & 2.2615\\
			0.011848 & 34.644 & 2.6923 & 2.2615 & 1.9538\\\hline
		\end{tabular}
	\end{center}
	\captionsetup{width=0.5\textwidth}\caption{Polytropic parameters of the first 3 EOSs with the smallest values of $e_i$ in the $4$-volume defined by Eq. \eqref{parametrosI}.}\label{tabla6}
\end{table}

\vspace{0.07cm}\begin{center} \textbf{Second realization of the test}\end{center}\vspace{0.07cm}

The results of the second realization of the test are shown in Table \ref{tabla7}.

\begin{table}[H]
	\begin{center}
		\begin{tabular}{| c | c | c | c | c |}\hline
			$e_i$ & $\log_{10} p_1$ & $\Gamma_1$ & $\Gamma_2$ & $\Gamma_3$\\\hline
			0.014579 & 34.644 & 2.5538 & 2.1077 & 2.8769\\
			0.015078 & 34.644 & 2.5538 & 2.1077 & 2.7231\\
			0.015809 & 34.644 & 2.5538 & 2.2615 & 1.9538\\\hline
		\end{tabular}
	\end{center}
	\captionsetup{width=0.5\textwidth}\caption{Polytropic parameters of the first 3 EOSs with the smallest values of $e_i$ in the $4$-volume defined by Eq. \eqref{parametrosI}.}\label{tabla7}
\end{table}

\vspace{0.07cm}\begin{center} \textbf{Third realization of the test}\end{center}\vspace{0.07cm}

The results of the third realization of the test are shown in Table \ref{tabla8}.

\begin{table}[H]
	\begin{center}
		\begin{tabular}{| c | c | c | c | c |}\hline
			$e_i$ & $\log_{10} p_1$ & $\Gamma_1$ & $\Gamma_2$ & $\Gamma_3$\\\hline
			0.0061922 & 34.644 & 2.6923 & 2.2615 & 1.9538\\
			0.012781 & 34.7 & 2.9692 & 1.8 & 3.0308\\
			0.012858 & 34.7 & 3.1077 & 1.8 & 3.0308\\\hline
		\end{tabular}
	\end{center}
	\captionsetup{width=0.5\textwidth}\caption{Polytropic parameters of the first 3 EOSs with the smallest values of $e_i$ in the $4$-volume defined by Eq. \eqref{parametrosI}.}\label{tabla8}
\end{table}

Comparing the results shown in Tables \ref{tabla6}, \ref{tabla7} and \ref{tabla8} with those shown in Table \ref{tabla1}, we conclude that $\Gamma_1$ and $\Gamma_3$ are the most affected polytropic parameters under a small variation of the original input data.

Note that the results presented here are just an estimation since we only show the first iteration of the meshing and refinement method. If several iterations were applied, we would expect the variations of $\Gamma_1$ and $\Gamma_3$ to be smaller.

\section{Conclusions}\label{conclusions}

The main objective of this paper was the development of a method to reconstruct the neutron star EOS from measurements of the mass and the $wI$-QNM spectra of different neutron stars. We have named it the piecewise polytropic meshing and refinement method since it starts with a wide mesh of polytropic parameters which is locally refined in the subsequent iterations. We have tested it considering the input data as $5$ GNH3 configurations ($5$ values of $M_{\text{GNH3}}$ and $\nu_{\text{GNH3}}$) and found that the algorithm reconstructs the EOS up to a given tolerance. The reconstructed EOS and the original GNH3 are very similar since the polytropic parameters of both EOS are similar itself. Moreover, the macroscopic parameters calculated from the reconstructed EOS are very similar to the ones calculated from the original GNH3 EOS. The meshing and refinement method has been applied for several EOSs and it works even better than for GNH3 (for instance, for APR4, SLy, ALF4). Hence, we are confident that the method would work efficiently with experimental data. Also, the algorithm is designed in such a way that it can reconstruct the EOS even if its polytopic parameters do not belong to the initial mesh.

As a subproduct of the meshing and refinement method, we developed some optimized algorithms based in the ones presented in (\cite{blazquez2013phenomenological},\cite{blazquez2014polar}) to calculate axial QNMs of thousands of EOSs in a reasonable time. To test the new algorithm, we calculated $w$-quasinormal modes of some realistic EOSs (APR4, SLy, GNH3, BHZBM, ALF4, BS3 and WSPHS) and plotted the results.

We also developed an optimized code to calculate slow rotation parameters together with the tidal deformation of thousands of EOS. We tested it by calculating the relation $\bar{\lambda}^{tid}_1(\bar{\lambda}^{tid}_2)$ for a binary system with $\mathcal{M}=1.188M_\odot$ for the different realistic EOSs considered here (namely APR4, SLy, GNH3, BHZBM, ALF4, BS3 and WSPHS) and showed that BS3 and WSPHS predict $\bar{\lambda}^{tid}$ values outside the $90\%$ confidence contour derived from the gravitational wave event GW170817 (for the low-spin scenario).

Let us now address the question of the application of the method to actual observations. The quasinormal modes of relativistic stars can be excited in several astrophysical processes, such as gravitational collapse to a neutron star, binary  coalescence leading to neutron star formation, pulsar ``glitches'', accretion of  matter or close encounter with other compact objects (\cite{kokkotas1999quasi},\cite{ferrari2008quasi}). In principle, in those situations gravitational waves would be produced and, eventually, could be detected. Which modes are more important in the gravitational radiation emitted in each situation depends on several factors, such as the amount of energy that can be stored in the mode, the presence of the other dissipative processes or the value of the frequency of the mode.

Numerical simulations of core collapse and binary coalescence suggest that the mode which would be most excited is the fundamental mode of the star (the $f$ mode). The $f$ mode has a typical frequency between $1.5$ and $3$ KHz, i.e., a region of the spectra where the current ground based detectors do not have sufficient sensitivity. The new generations of interferometric detectors will have better sensitivity in the high frequency region and, in principle, it could be  possible to detect the $f$ mode of neutron stars (\cite{aligo+},\cite{torres2019observing},\cite{hild2008pushing},\cite{dooley2016geo}).

In this paper we use axial $w$ modes, since numerical calculations for axial $w$ modes are simpler than those for polar $f$ modes. However, the piecewise polytropic meshing and refinement method can be applied with $f$ modes and it is now under development.

The typical frequency of the fundamental axial $w$ mode of a 
relativistic star is over $5$ KHz, which makes it more unlikely to be detected by the current ground detectors than the $f$ mode. Anyway, let us mention that recently it has been shown that these modes can be excited in the collapse of a neutron star to a black hole. The $w$ modes of the neutron star are excited soon before the black hole formation and the gravitational radiation of the process contains $w$ modes of the collapsing star and the born black hole \cite{baiotti2005gravitational} (note that, in this situation, the EOS would contain more microphysics information). Others studies of the excitability of the $w$ modes can be found in (\cite{andersson1996gravitational},\cite{ferrari1999stellar},\cite{andrade1999excitation},\cite{tominaga1999gravitational},\cite{bernuzzi2008dynamical}).

In conclusion, in order to make gravitational astereoseismology we need gravitational wave detectors more sensitive to high frequencies. We hope that, in the next generations of the detectors, it will be possible to detect quasinormal modes of relativistic stars \cite{torres2019observing}. Also, we expect that the algorithm developed in this paper (the piecewise polytropic meshing and refinement method), which can be applied with different global properties of the star ($w$ modes in this paper, $f$ modes,\dots), will be useful to study the EOS and the properties of neutron stars.

\section{Acknowledgments}

We would like to thank J. L. Blázquez Salcedo and F. Navarro Lérida for their comments, suggestions and very helpful discussions.

\onecolumngrid

\appendix

\section{Slowly rotating relativistic stars}\label{appendixA}

We here will summarize the equations for uniformly slowly rotating relativistic stars, following Ref. \cite{hartle1967slowly}, in the International System of Units. Coordinates can be chosen so that the line element has the form
\begin{myequation}\label{metrica_slow_rotation}
ds^2=-H^2(r,\theta)(cdt)^2+Q^2(r,\theta)dr^2+r^2K^2(r,\theta)\left\{d\theta^2+\sin^2\theta\left[d\varphi-L(r,\theta)dt\right]^2\right\}.
\end{myequation}

The fluid's four-velocity will be considered circular, which means that $u^\mu\propto \delta^\mu_t+\frac{\Omega}{c}\delta^\mu_\varphi$, where $\Omega=\text{constant}$ is the angular velocity of the fluid. From the normalization condition $u_\mu u^\mu=-c^2$ it follows that
\begin{myequation}
	u^\mu=\left[-\left(g_{tt}+2\frac{\Omega}{c} g_{t\varphi}+\frac{\Omega^2}{c^2}g_{\varphi\varphi}\right)\right]^{-1/2}(c,0,0,\Omega).
\end{myequation}

The metric functions $H^2$, $Q^2$ and $K^2$ must be even functions of $\Omega$, and $L$ must be an odd function of $\Omega$ \cite{hartle1967slowly}. Hence, these functions can be expanded in power series of $\varepsilon\propto\Omega$ as follows
\begin{subequations}\label{eq2}
	\begin{gather}
	\begin{align}
	&H^2(r,\theta)=e^{\nu(r)}[1+2h(r,\theta)+\mathcal{O}(\varepsilon^4)],\\
	&Q^2(r,\theta)=e^{\lambda(r)}\left[1+\frac{2G}{c^2}\frac{e^{\lambda(r)}}{r}m(r,\theta)+\mathcal{O}(\varepsilon^4)\right],\\
	&K^2(r,\theta)=1+2k(r,\theta)+\mathcal{O}(\varepsilon^4),\\
	&L(r,\theta)=\omega(r,\theta)+\mathcal{O}(\varepsilon^3).
	\end{align}
	\end{gather}
\end{subequations}

\subsection{First order slow rotation}

From the perturbed Einstein equation $\delta G^{(1)}_{t\varphi}=\frac{8\pi G}{c^4}\delta T^{(1)}_{t\varphi}$ it follows that $\omega(r,\theta)=\omega(r)$ \cite{hartle1967slowly}. The resulting Einstein equation for $\bar{\omega}(r)=\Omega-\omega(r)$ can be written as
\begin{myequation}\label{orden1}
	\frac{1}{r^4}\frac{d}{dr}\left(r^4j\frac{d\bar{\omega}}{dr}\right)+4r^{-1}\frac{dj}{dr}\bar{\omega}=0,
\end{myequation}
where
\begin{myequation}
	j=e^{-\frac{\nu+\lambda}{2}}.
\end{myequation}

\vspace{0.07cm}\begin{center}\textbf{- Exterior solution to the} $\boldsymbol{\mathcal{O}(\varepsilon)}$ \textbf{equation}\end{center}\vspace{0.07cm}

In the exterior of the star we know that $\nu=-\lambda$, and hence the solution to equation \eqref{orden1} is given by
\begin{myequation}
	\bar{\omega}(r)=\Omega-\frac{2GJ}{c^2r^3},
\end{myequation}
where $J$ is the total angular momentum of the star,
\begin{myequation}
	J=\frac{c^2R^4}{6G}\frac{d\bar{\omega}}{dr}\bigg\rvert_{r=R},
\end{myequation}
as defined in Ref. \cite{hartle1967slowly}. The moment of inertia can be calculated as
\begin{myequation}
	I=\frac{J}{\Omega}.
\end{myequation}

\subsection{Second order slow rotation}

If an expansion in spherical harmonics of the metric \eqref{metrica_slow_rotation} is made, following Ref. \cite{regge1957stability}, one finds that it takes the form
\begin{subequations}
	\begin{gather}
	\begin{align}
	&h(r,\theta)=h_0(r)+h_2(r)P_2(\cos\theta)+\dots,\\
	&m(r,\theta)=m_0(r)+m_2(r)P_2(\cos\theta)+\dots,\\
	&k(r,\theta)=k_2(r)P_2(\cos\theta)+\dots,
	\end{align}
	\end{gather}
\end{subequations}
because $H^2$, $Q^2$ and $K^2$ transform like scalars under rotations \cite{hartle1967slowly}. At this order, the border of the star is deformed,
\begin{myequation}
	r\to r+\xi(r,\theta)+\mathcal{O}(\varepsilon^4).
\end{myequation}
Second order slow rotation also perturbs the pressure and the energy density,
\begin{myequation}
	p\to p-\xi(r,\theta)\frac{dp}{dr}+\mathcal{O}(\varepsilon^4),\quad \epsilon\to \epsilon-\xi(r,\theta)\frac{d\epsilon}{dr}+\mathcal{O}(\varepsilon^4).
\end{myequation}
The function $\xi(r,\theta)$ is proportional to $\Omega^2$ and transforms like a scalar under rotations \cite{hartle1967slowly},
\begin{myequation}
	\xi(r,\theta)=\xi_0(r)+\xi_2(r)P_2(\cos\theta)+\dots.
\end{myequation}
It can be shown that all the coefficients with $l>2$ in the expansions of $h$, $m$, $k$ and $\xi$ must vanish \cite{hartle1967slowly}. The reduction in the number of values of $l$ from infinity to $2$ in the slow rotation approximation greatly simplifies the calculations.

Following Ref. \cite{hartle1967slowly}, we define a function $p^*(r,\theta)$ such that 
\begin{myequation}
	\xi(r,\theta)=-c^2\left(\epsilon+\frac{p}{c^2}\right)\left(\frac{dp}{dr}\right)^{-1}p^*(r,\theta).
\end{myequation}

From the conservation of the stress-energy tensor, $\nabla_\nu T^{\theta\nu}=0$, one finds that
\begin{myequation}
	p_2^*=-h_2-\frac{r^2\bar{\omega}^2e^{-\nu}}{3c^2}.
\end{myequation}
The previous relation and
\begin{myequation}
	p_0^*=\text{constant}-h_0+\frac{r^2\bar{\omega}^2e^{-\nu}}{3c^2},
\end{myequation}
which is also obtained in Ref. \cite{hartle1967slowly}, represent two first integrals of motion. 

The equations for different values of $l$ are not coupled together, and can then be considered separately \cite{hartle1967slowly}.

\subsubsection{The $l=0$ equations}

From the perturbed Einstein equations $\delta G_{tt}^{(2)}=\frac{8\pi G}{c^4}\delta T_{tt}^{(2)}$ and $\delta G_{rr}^{(2)}=\frac{8\pi G}{c^4}\delta T_{rr}^{(2)}$ it follows, respectively, that
\begin{small}
	\begin{subequations}\label{orden2_l0}
		\begin{gather}
		\begin{align}
		&\frac{dm_0}{dr}=\frac{r^4j^2\bar{\omega}'^2}{12G}-\frac{r^3(j^2)'\bar{\omega}^2}{3G}+4\pi c^2r^2\left(\epsilon+\frac{p}{c^2}\right)\frac{d\epsilon}{dp}p_0^*,\\
		&\frac{dp_0^*}{dr}=\frac{d}{dr}\left(\frac{r^2\bar{\omega}^2e^{-\nu}}{3c^2}\right)-\frac{1}{1-\frac{2Gm}{c^2r}}\left[\frac{Gm_0}{c^2r}\left(\frac{1}{r}+\nu'\right)-\frac{1}{12c^2}j^2r^3\bar{\omega}'^2+\frac{4\pi G}{c^2}r\left(\epsilon+\frac{p}{c^2}\right)p_0^*\right].
		\end{align}
		\end{gather}
	\end{subequations}
\end{small}

\subsubsection{The $l=2$ equations}

Combining the $(\theta,\theta)$ and the $(\varphi,\varphi)$ components as $\delta G_{\varphi\varphi}^{(2)}-\frac{8\pi G}{c^4}\delta T_{\varphi\varphi}^{(2)}=\sin^2\theta\left[\delta G_{\theta\theta}^{(2)}-\frac{8\pi G}{c^4}\delta T_{\theta\theta}^{(2)}\right]$, one finds a first integral of motion given by
\begin{myequation}
	h_2+\frac{Gm_2}{c^2r\left(1-\frac{2Gm}{c^2r}\right)}=\frac{r^4\bar{\omega}'^2j^2}{6c^2}-\frac{(j^2)'r^3\bar{\omega}^2}{3c^2}.
\end{myequation}
Finally, from components $(r,r)$ and $(r,\theta)$ one finds the system of equations
\begin{small}
	\begin{subequations}\label{orden2_l2}
		\begin{gather}
		\begin{align}
		&\frac{dv}{dr}=-h_2\nu'+\left(\frac{1}{r}+\frac{\nu'}{2}\right)\left[\frac{r^4j^2\bar{\omega}'^2}{6c^2}-\frac{r^3(j^2)'\bar{\omega}^2}{3c^2}\right],\\
		&\frac{dh_2}{dr}=\left\{-\nu'+\frac{1}{c^2\left(1-\frac{2Gm}{c^2r}\right)\nu'}\left[8\pi G\left(\epsilon+\frac{p}{c^2}\right)-\frac{4Gm}{r^3}\right]\right\}h_2-\frac{4v}{r^2\left(1-\frac{2Gm}{c^2r}\right)\nu'}\nonumber\\
		&\qquad+\frac{1}{6}\left[\frac{1}{2}\nu'r-\frac{1}{r\left(1-\frac{2Gm}{c^2r}\right)\nu'}\right]\frac{r^3j^2\bar{\omega}'^2}{c^2}-\frac{1}{3}\left[\frac{1}{2}\nu'r+\frac{1}{r\left(1-\frac{2Gm}{c^2r}\right)\nu'}\right]\frac{r^2(j^2)'\bar{\omega}^2}{c^2},
		\end{align}
		\end{gather}
	\end{subequations}
\end{small}
where $v=h_2+k_2$.

\setcounter{subsubsection}{0}

\vspace{0.07cm}\begin{center}\textbf{- Exterior solution to the} $\boldsymbol{\mathcal{O}(\varepsilon^2)}$ \textbf{equations}\end{center}\vspace{0.07cm}

\subsubsection{The $l=0$ equations}

The solution of the system of equations \eqref{orden2_l0} (in terms of $h_0$ instead of $p_0^*$) in the exterior of the star is given by
\begin{small}
	\begin{subequations}
		\begin{gather}
		\begin{align}
		&m_0=\delta M-\frac{GJ^2}{c^4r^3},\\
		&h_0=\frac{G\left(-\delta M+\frac{GJ^2}{c^4r^3}\right)}{c^2r\left(1-\frac{2GM}{c^2r}\right)},
		\end{align}
		\end{gather}
	\end{subequations}
\end{small}
where $\delta M$ is the change in mass of the perturbed configuration from its non perturbed value.

\subsubsection{The $l=2$ equations}

The solution of the system of equations \eqref{orden2_l2} in the exterior of the star is given by

\begin{small}
	\begin{subequations}
		\begin{gather}
		\begin{align}
		&v=KQ_2^1(\zeta)\frac{2GM}{c^2r}\frac{1}{\sqrt{1-\frac{2GM}{c^2r}}}-\frac{G^2J^2}{c^6r^4},\\
		&h_2=KQ_2^2(\zeta)+\frac{G^2J^2}{c^6r^4}\left(\frac{c^2r}{GM}+1\right),
		\end{align}
		\end{gather}
	\end{subequations}
\end{small}
where $\zeta=\frac{c^2r}{GM}-1$ and $Q_l^m(\zeta)$ is the associated Legendre function of the second kind \cite{hartle1967slowly}. The constant $K$ is related to the mass quadrupole moment of the configuration by the relation \cite{hartle1967slowly}
\begin{myequation}
	Q=\frac{J^2}{Mc^2}+\frac{8KG^2M^3}{5c^4}.
\end{myequation}

\section{Tidally deformed neutron stars}\label{appendixB}

In this section, we will consider neutron stars in binary systems. The primary star will be modeled as a non-rotating star that is tidally deformed by its companion, the secondary star \cite{yagi2013love}. The deformation due to tidal effects will be implemented via perturbations over static and spherically symmetric spacetime, as we did when introducing slow rotation and the QNM approximation.

As we are interested in a non-rotating tidally deformed neutron star, we set $\Omega=0$. It can be shown that the metric perturbation for the second order tidal deformation is the same as the $l=2$ part of the second order slow rotation perturbation. Setting $\bar{\omega}=0$ in the system of ordinary differential equations given by Eq. \eqref{orden2_l2}, 
\begin{small}
	\begin{subequations}
		\begin{gather}
		\begin{align}
		&\frac{dv^H}{dr}=-h_2^H\nu',\\
		&\frac{dh_2^H}{dr}=\left\{-\nu'+\frac{1}{c^2\left(1-\frac{2Gm}{rc^2}\right)\nu'}\left[8\pi G\left(\epsilon+\frac{p}{c^2}\right)-\frac{4Gm}{r^3}\right]\right\}h_2^H-\frac{4v^H}{r^2\left(1-\frac{2Gm}{rc^2}\right)\nu'}.
		\end{align}
		\end{gather}
	\end{subequations}
\end{small}
The superscript $H$ stands for the fact that these equations are simply the homogeneous part of the full system of equations \eqref{orden2_l2}. Once we have solved these equations numerically, we calculate the tidal Love number, $k_2^{tid}$. The tidal Love number is related to how easy or difficult it would be to deform a star. It is given by \cite{hinderer2008tidal}

\begin{myequation}
	\begin{array}{ll}
		k_2^{tid}=\frac{8C^5}{5}(1-2C)^2\left[2+2C(y-1)-y\right]\left\{2C\left[6-3y\right.\right.\\\\
		\left.+3C(5y-8)\right]+4C^3\left[13-11y+C(3y-2)+2C^2(1+y)\right]\\\\
		\left.+3(1-2C)^2\left[2-y+2C(y-1)\right]\log(1-2C)\right\}^{-1},
	\end{array}
\end{myequation}
where
\begin{myequation}
	C=\frac{GM}{c^2R},\quad y=R\left[\frac{1}{h_2^H}\frac{dh_2^H}{dr}\right]\bigg\rvert_{r=R}-\frac{4\pi R^3\epsilon_{sup}}{M}.
\end{myequation}
$C$ is known as the compactness parameter, and $\epsilon_{sup}$ is the energy density$/c^2$ at the surface of the star, if non-zero \cite{hinderer2010tidal}. We will be interested in calculating the so-called tidal Love parameter, which is given in terms of the tidal Love number $k_2^{tid}$ and the compactness parameter $C$ as
\begin{myequation}
	\bar{\lambda}^{tid}=\frac{2k_2^{tid}}{3C^5}.
\end{myequation}

\subsection{I-Love-Q universal relations}

Neutron stars are not only characterized by their mass and radius. They are also characterized by how fast they can spin given a fixed angular momentum, through their moment of inertia, how much they can be deformed away from sphericity, through their quadrupole moment, and how easy or difficult it would be to deform them, through their tidal Love parameter \cite{yagi2013love}. All these quantities depend sensitively on the star's internal structure, and hence on unknown nuclear physics. 

Following Ref. \cite{yagi2013love}, let us define the adimensional quantities
\begin{myequation}\label{IQparameters}
	\bar{I}=\frac{c^4I}{G^2M^3}\quad\text{and}\quad\bar{Q}=\frac{c^2MQ}{J^2}.
\end{myequation}
The I-Love-Q universal relations are relations between the moment of inertia ($\bar{I}$), the quadrupole moment ($\bar{Q}$) and the tidal Love parameter ($\bar{\lambda}^{tid}$) that do not depend on the internal structure of neutron stars (i.e., do not depend on the EOS of the star). This can be used in our benefit to obtain any of these parameters measuring only one of them.

\bibliographystyle{unsrt}
\bibliography{Inverse_QNM}

\end{document}